\documentclass[10pt,final,journal,letterpaper,oneside,twocolumn]{IEEEtran}
\usepackage[utf8]{inputenc}

\usepackage{amsmath}
\usepackage{amsthm}
\usepackage{amssymb}
\usepackage{mathtools}
\usepackage{bm, bbm}
\usepackage[scr=boondoxo,scrscaled=1.05]{mathalfa}
\usepackage{enumitem}

\usepackage{caption,subcaption}

\usepackage{cite}   
\usepackage{xcolor}
\usepackage{comment}
\usepackage{lipsum}
\usepackage{booktabs}

\usepackage[hidelinks,draft=false]{hyperref}

\DeclarePairedDelimiterXPP{\KL}[2]{D_\textnormal{KL}}{(}{)}{}{%
#1\:\delimsize\|\:#2%
}
\DeclarePairedDelimiterXPP{\RD}[2]{D_{\alpha}}{(}{)}{}{%
#1\:\delimsize\|\:#2%
}
\DeclarePairedDelimiterXPP\Prob[1]{\mathbb{P}}{\lbrace}{\rbrace}{}{

#1}
\DeclarePairedDelimiterXPP{\lnorm}[2]{}{\lVert}{\rVert}{_{#2}}{#1}

\newcommand{\bA}{\ensuremath{\mathbb{A}}}
\newcommand{\bE}{\ensuremath{\mathbb{E}}}
\newcommand{\bR}{\ensuremath{\mathbb{R}}}

\newcommand{\cS}{\ensuremath{\mathcal{S}}}
\newcommand{\cT}{\ensuremath{\mathcal{T}}}
\newcommand{\cW}{\ensuremath{\mathcal{W}}}
\newcommand{\cV}{\ensuremath{\mathcal{V}}}
\newcommand{\cX}{\ensuremath{\mathcal{X}}}
\newcommand{\cY}{\ensuremath{\mathcal{Y}}}
\newcommand{\cZ}{\ensuremath{\mathcal{Z}}}

\newcommand{\stack}[2]{\stackrel{\mathclap{(#1)}}{#2}}

\newtheoremstyle{mytheoremstyle} 
    {\topsep}                    
    {\topsep}                    
    {\itshape}                   
    {}                           
    {\bf}                   
    {.}                          
    {.5em}                       
    {}  

\theoremstyle{mytheoremstyle}

\newtheorem{lemma}{Lemma}
\newtheorem{claim}{Claim}
\newtheorem{proposition}{Proposition}
\newtheorem{definition}{Definition}

\newtheorem{theorem}{Theorem}

\newtheorem{remark}{Remark}
\newtheorem{assumption}{Assumption}

\title{Upper Bounds on the Generalization Error of Private Algorithms {for Discrete Data}}
\author{Borja~Rodr\'iguez-G\'alvez,~\IEEEmembership{Graduate~Student~Member,~IEEE}, Germ\'an~Bassi,~\IEEEmembership{Member,~IEEE}, and Mikael~Skoglund,~\IEEEmembership{Fellow,~IEEE}%
\thanks{
Manuscript received November 12, 2020; revised May 26, 2021; accepted August 27, 2021. This work was supported in part by the Knut and Alice Wallenberg Foundation, the Swedish Foundation for Strategic Research, and the Swedish Research Council {under contract 2019-03606} \textit{(Corresponding author: Borja Rodr\'iguez-G\'alvez)}.

Borja~Rodr\'iguez-G\'alvez and Mikael~Skoglund are with the Division of Information Science and Engineering, Department of Intelligent Systems, KTH Royal Institute of Technology, 114 28 Stockholm, Sweden (e-mail: borjarg@kth.se; skoglund@kth.se).

Germ\'an~Bassi was with KTH Royal Institute of Technology 114 28 Stockholm, Sweden. He is now with Ericsson Research, 164 83 Stockholm, Sweden (e-mail: german.bassi@ericsson.se).

Communicated by V. Y. F. Tan, Associate Editor for Machine Learning.
Digital Object Identifier 10.1109/TIT.2021.3111480

Copyright (c) 2017 IEEE. Personal use of this material is permitted.  However, permission to use this material for any other purposes must be obtained from the IEEE by sending a request to pubs-permissions@ieee.org.}
}

\begin{document}

\maketitle

\begin{abstract}
In this work, we study the generalization capability of algorithms from an information-theoretic perspective.
It has been shown that the expected generalization error of an algorithm is bounded from above by a function of the relative entropy between the conditional probability distribution of the algorithm's output hypothesis, given the dataset with which it was trained, and its marginal probability distribution.
We build upon this fact and introduce a mathematical formulation to obtain upper bounds on this relative entropy.
{Assuming that the data is discrete,} we then develop a strategy using this formulation, based on the method of types and typicality, to find explicit upper bounds on the generalization error of stable algorithms, i.e., algorithms that produce similar output hypotheses given similar input datasets.
In particular, we show the bounds obtained with this strategy for the case of $\epsilon$-DP and $\mu$-GDP algorithms.
\end{abstract}


\section{Introduction}
\label{sec:introduction}

\IEEEPARstart{A}{learning} algorithm is a mechanism that takes a collection of data samples as an input and outputs a hypothesis.
The usage of this type of algorithm spans from estimating the sinusoidal parameters of a received, noisy signal~\cite{stoica1989maximum} to detecting and localizing a tumor from an MRI scan~\cite{havaei2017brain}.
The generalization capability of a learning algorithm indicates its ability to perform similarly in new, unseen data, as it performed in the finite amount of data with which it was trained.
Therefore, characterizing this capability allows us to evaluate the worth of an algorithm outside of the training data and, with a proper characterization framework, design robust algorithms.

A common characterization of the generalization capability of a learning algorithm is given by bounding its \emph{generalization error}.
This error is defined as the difference between the performance of an algorithm's output hypothesis evaluated on the total population of some distribution and evaluated on a finite set of $N$ samples of the said distribution.
A recent line of work has established that an upper bound on the expected value of this error is controlled by the mutual information between the dataset the algorithm is trained on and the hypothesis it produces~\cite{russo_2020_bias, xu2017information, jiao_2017_dependence, bu_2019_tightening}; that is, the 
{less} an algorithm relies on the training data to produce its hypothesis, the 
{more} it generalizes. 

In the context of statistical analysis, a privacy mechanism is an algorithm that answers queries about a given dataset in a way that is informative about the queries themselves but not the specific records (or samples) in the dataset.
Private learning algorithms are a special class of learning algorithms that employ a privacy mechanism to analyze the training dataset; hence, these algorithms produce hypotheses that are uninformative about the dataset with which they were trained.
Therefore, there is a connection between these two desirable properties: privacy and generalization.

This connection has already been studied for differentially private algorithms, e.g., \cite{dwork_generalization_2015, dwork_preserving_2015}.
\emph{Differential privacy}~\cite{dwork_dp_2014} was the first mathematically rigorous framework for privacy.
In its most basic form, it defines the privacy level of an algorithm as the maximum log-likelihood ratio between the algorithm's outputs for two similar inputs.
If this ratio is bounded for any two similar inputs, an adversary cannot easily distinguish the said inputs.
Hence, if an adversary cannot determine which of two inputs of an algorithm is the cause of the observed output, it means that the distribution of outputs does not depend to a great extent on any one of the input samples.

Recently, a new definition of privacy was introduced in~\cite{dong2019gaussian}: \emph{$f$-differential privacy} ($f$-DP). This privacy framework is described in terms of a hypothesis test to distinguish between two inputs of an algorithm based on the output the algorithm produced with one of them.
More specifically, an algorithm is said to be $f$-DP if the trade-off curve described by the Type-I and Type-II errors of the aforementioned hypothesis test lies above the non-increasing, convex function $f$~\cite{dong2019gaussian}. 
The function $f$ is thus the parameter that determines the privacy level of the algorithm.
A particular, and important, parametrization of these algorithms is given by \emph{$\mu$-Gaussian differentially private} ($\mu$-GDP) algorithms, where $f$ is defined as the trade-off function between two unit-variance Gaussian distributions with mean $0$ and $\mu$, respectively. 

When used in a learning setting, the aforementioned private algorithms share an important property: stability or smoothness.
This means that similar input datasets produce similar output hypotheses.
In this work, we focus on studying the generalization error of learning algorithms that share this property.

\subsection{Related work}

Traditional approaches to characterizing the generalization error are based on the Rademacher complexity, the Vapnik--Chervonenkis (VC) dimension, probably approximately correct (PAC) Bayesian theory, compression bounds, or stability theory, to name a few.
The reader is referred to~\cite{shalev2014understanding} for a thorough exposition of these topics.

More recently, there have been several approaches that leverage information-theoretic measures in order to bound the generalization error.
For instance, the authors of~\cite{xu2017information} and~\cite{asadi2018chaining} use the mutual information between the (entire) training dataset and the output hypothesis, while the authors of~\cite{bu_2019_tightening} follow up this work using the mutual information between a single data sample and the output hypothesis.
The use of the mutual information, the lautum information, and the maximal leakage is proposed in~\cite{issa2019strengthened} while the authors of~\cite{wang2019information} leverage the Wasserstein distance.

In this work, we follow the path started by Russo and Zou in~\cite{russo_2020_bias}, where they introduce a framework for characterizing the bias in data exploration.
More precisely, they show that the expected bias of an estimator is bounded from above by a function of the mutual information between the estimator and the rule for selecting such {an} estimator.
Using this framework, the authors of~\cite{xu2017information} show that this is equivalent to saying that the generalization error is bounded from above by a function of the mutual information between the dataset and the output hypothesis of a learning algorithm.
It is important to note that, in~\cite{russo_2020_bias}, the bias is assumed to be sub-Gaussian, which implies that the risk function is sub-Gaussian in~\cite{xu2017information}.
However, in~\cite{jiao_2017_dependence}, it is shown that, if the bias is sub-Gamma or sub-Exponential, then the bias is also bounded from above by a function (albeit different) of the mutual information between the estimator and the rule.

All these works show that the generalization capability of an algorithm is tied with a measure of the dependence between the output hypothesis and the training dataset, which is in line with the bounds based on stability theory, e.g., \cite[Chapter 13]{shalev2014understanding}, \cite{bousquet2002stability, raginsky2016information, bassily_algorithmic_2016, feldman2019high}.
In the case of differentially private algorithms, this property has already been leveraged (i) to study the accuracy of an algorithm's answers to an analyst's queries, e.g., with the following concentration bounds~\cite[Theorems~9--11]{dwork_preserving_2015}, \cite[Theorem~7.2]{bassily_algorithmic_2016}, and~\cite[Lemmas~7, 15, and~22]{jung_new_2020}; and (ii) to find bounds on the (approximate) max-information between the input dataset and an algorithm's output hypothesis~\cite[Theorems~7 and~8]{dwork_generalization_2015}.

\subsection{Contribution and organization of the paper}

In this paper, we introduce a mathematical formulation to bound from above the relative entropy between the conditional distribution of the output hypothesis of a learning algorithm (given the input dataset) and its marginal distribution, thus bounding the generalization error as well.
This formulation is based on a variational approximation, in the form of a mixture, of the marginal distribution of the output hypothesis of an algorithm.
It leverages the trade-off between the responsibilities (weights) of the mixture elements, and the similarity between said elements and the distribution of the hypothesis, given a particular dataset.

If an algorithm is stable, i.e., similar input datasets produce similar hypotheses, then the aforementioned trade-off can be exploited to find tighter bounds on the generalization error.
This is done by selecting the mixture elements to be the hypotheses' conditional distributions that best cover the set of possible datasets.
The rule for selecting such a cover of the hypothesis space is determined by the level of stability of the algorithm.

We develop a strategy, based on the method of types, to find explicit upper bounds on the relative entropy using this formulation {when the data is discrete}.
First, we show that this strategy can recover a trivial bound on the mutual information. 
Then, we leverage the stability property of $\epsilon$-DP and $\mu$-GDP algorithms to obtain tighter bounds. 
These results, combined with~\cite[Theorem~1]{xu2017information}, allow us to obtain upper bounds on the expected generalization error of (private) learning algorithms.
We present (a summary of) these bounds in the following two theorems; the terminology and the proper definition of the quantities in the statement of the theorems are introduced in Section~\ref{sec:problem_formulation}.

\begin{theorem}
\label{th:main_result_general}
Let $S \in \cS$ be a dataset of $N$ instances $Z_n \in \cZ$, $|\cZ| < \infty$, sampled i.i.d. from $P_Z$.
Let also $W \in \cW$ be a hypothesis obtained with an algorithm $\bA$, characterized by $P_{W|S}$, such that $\bA(S) \stackrel{\textnormal{d}}{=} \bA(\mathcal{P}(Z^N))$ for any permutation $\mathcal{P}$ of any sequence $Z^N$ obtained by enumerating $S$.
Suppose the non-negative loss function $\ell(w,Z)$ is $\sigma$-sub-Gaussian under $P_Z$ for all $w \in \cW$.
Then,
\begin{equation}
    \big| \bE_{P_{W,S}} [\textnormal{gen}(W,S)] \big| \leq \sqrt{2 \sigma^2 (|\cZ| -1) \frac{\log(N+1)}{N}}.
\end{equation}
\end{theorem}

\begin{theorem}
\label{th:main_result_private}
Let $S \in \cS$ be a dataset of $N$ instances $Z_n \in \cZ$, $|\cZ| < \infty$, sampled i.i.d. from $P_Z$.
Let also $W \in \cW$ be a hypothesis obtained with an algorithm $\bA$, characterized by $P_{W|S}$, such that $\bA(S) \stackrel{\textnormal{d}}{=} \bA(\mathcal{P}(Z^N))$ for any permutation $\mathcal{P}$ of any sequence $Z^N$ obtained by enumerating $S$.
Suppose the non-negative loss function $\ell(w,Z)$ is $\sigma$-sub-Gaussian under $P_Z$ for all $w \in \cW$.
Then, if $\bA$ is $\gamma$-DP with $\gamma \leq 2$ or $\gamma$-GDP with $\gamma \leq 2/\sqrt{|\cZ|}$,
\begin{equation}
    \big| \bE_{P_{W,S}} [\textnormal{gen}(W,S)] \big| \preceq \sqrt{2 \sigma^2 |\cZ| \frac{\log(\gamma \sqrt{N \log N})}{N}},
    \label{eq:main_result_private}
\end{equation}
where the meaning of $\preceq$ is introduced in Definition~\ref{def:preceq}.
\end{theorem}

The paper is organized as follows.
First, in Section~\ref{sec:notation}, we introduce the notation used throughout the paper.
Then, in Sections~\ref{sec:problem_formulation} and~\ref{sec:preliminaries}, we mathematically define the problem and introduce some theory relevant for developing our findings.
Section~\ref{sec:main_results} presents our main results on strict upper bounds for the relative entropy, and thus the expected generalization error as well.
Finally, in Section~\ref{sec:discussion} we summarize and contextualize our findings.
All proofs can be found in the Appendices.

\section{Notation}
\label{sec:notation}

In this section, we introduce the notation and definitions that will be used throughout the article.

Calligraphic letters (e.g., $\cX$) represent sets; except for the special sets of the real, positive real, and positive integer numbers, which are denoted by $\bR$, $\bR_+$, and $\mathbb{Z}_+$, respectively.
Capital letters (e.g., $X$) mostly denote random variables, while their lower-case (e.g., $x$) and script-styled (e.g., $\mathscr{X}$) counterparts denote instances of the said random variables and their sigma algebras, respectively.
Unless stated otherwise, the same letter is used for a random variable, its instances, and the set it is contained in.


A random variable $X$ is a function $X: \Omega \rightarrow \cX$ from the probability space $(\Omega, \mathcal{F}, \mathbb{P})$ to the target space $(\cX, \mathscr{X})$.
It is described by its probability distribution $P_X: \mathscr{X} \rightarrow [0,1]$, which is defined as the set function $P_X(B) = \mathbb{P} \lbrace X \in B \rbrace$, for all $B \in \mathscr{X}$.
We write $P_X(x)$ as a shorthand to refer to $P_X(\lbrace x \rbrace)$. 

If we consider more than one random variable, e.g., $X$ and $Y$, we write their joint probability distribution as $P_{X,Y}: \mathscr{X} \otimes \mathscr{Y} \rightarrow [0,1]$ and their product distribution as $P_{X} \times P_{Y}: \mathscr{X} \otimes \mathscr{Y} \rightarrow [0,1]$. 
Moreover, we also write the conditional distribution of $Y$ given $X$ as $P_{Y|X} : \mathscr{Y} \times \cX \rightarrow [0,1]$, which defines a probability distribution $P_{Y|X=x}$ over $(\mathcal{Y},\mathscr{Y})$ for each element $x \in \cX$.
Finally, we abuse notation and write $P_{Y|X} \times P_X = P_{X,Y}$ since $P_{X,Y}(B) =  \int \big( \int \chi_B \big((x,y) \big) dP_{Y|X=x}(y) \big) dP_X(x)$ for all $B \in \mathscr{X} \otimes \mathscr{Y}$, where $\chi_B$ is the characteristic function of the set $B$.

For a function $\varphi: \cX \rightarrow \cY$, the expected value of $\varphi(X)$ is denoted as $\mathbb{E}_{x \sim P_X}[\varphi(x)]$ or $\mathbb{E}_{P_X}[\varphi(X)]$. We may use either $\mathbb{E}_{P_X}[\varphi]$ or $\mathbb{E}[\varphi(X)]$ as shorthands when it is clear from context. 
The function $\log$ denotes the natural logarithm.

\begin{definition}
Let $P$ and $Q$ be two probability distributions.
Then, if $P$ is absolutely continuous with respect to $Q$, i.e., $P\ll Q$, the \emph{relative entropy} or \emph{Kullback--Leibler (KL) divergence} of $P$ from $Q$ is defined as
\begin{equation}
	\KL{P}{Q}  \triangleq \mathbb{E}_{P} \bigg[ \log \left( \frac{dP}{dQ} \right) \bigg],
\end{equation}
where $dP/dQ$ is the Radon--Nikodym derivative.
Otherwise $\KL{P}{Q} = +\infty$.
\end{definition}

\begin{definition}
Let $X$ and $Y$ be two random variables with probability distributions $P_X$ and $P_Y$, and with joint probability distribution $P_{X,Y}$.
Then, the \emph{mutual information} between $X$ and $Y$ is defined as 
\begin{equation}
	I(X;Y) \triangleq \KL[\big]{P_{X,Y}}{P_X \times P_Y}.
\end{equation}
\end{definition}

\begin{definition}
\label{def:preceq}
We say that a function $f: \mathcal{X} \times \mathbb{Z}_+ \rightarrow \mathbb{R}$ is \emph{asymptotically bounded from above} by $g: \mathcal{X} \times \mathbb{Z}_+ \rightarrow \mathbb{R}$ if there exist $k \in \mathbb{R}_+$ and $n_0 \in \mathbb{Z}_+$ such that for all $n > n_0$ and $x \in \mathcal{X}$, we have that $f(x,n) \leq k g(x,n)$. 
We denote this by $f \preceq g$.
\end{definition}

\section{Problem Formulation}
\label{sec:problem_formulation}

A \emph{dataset} $S \in \cS$ is a collection of $N$ \emph{instances} $Z_n\in\cZ$, i.e., $S = \lbrace Z_1, Z_2, \ldots, Z_N \rbrace$.
A \emph{learning algorithm} $\mathbb{A}: \cS \rightarrow \cW$, characterized by the conditional probability {distribution $P_{W|S}$}, is a mechanism that takes a dataset of instances $S$ as input and returns as output a hypothesis {$w \in \cW$ sampled from $P_{W|S}$}.

\begin{assumption}
\label{ass:samples}
We consider that the instances are sampled i.i.d. from $P_Z$; therefore, $S$ is a random variable with probability distribution $P_S = P_Z^{\times N}$ over $\cZ^{N}$.
Furthermore, we assume that $|\cZ|<\infty$.
\end{assumption}

\begin{assumption}
\label{ass:algorithm}
We consider algorithms that operate on a \emph{set} $S = \lbrace Z_1, Z_2, \ldots, Z_N \rbrace$ of $N$ instances $Z_n \in \cZ$.
This means that the hypothesis generated by the algorithm $\bA$ with the dataset $S$ has the same distribution as the one generated by $\bA$ with a permutation $\mathcal{P}(Z^N)$ of any sequence $Z^N = (Z_1, Z_2, \ldots, Z_N)$ generated by enumerating $S$; i.e., $\bA(S) \stackrel{\textnormal{d}}{=} \bA(\mathcal{P}(Z^N))$ for all permutations $\mathcal{P}$, where the equality should be understood as ``in distribution.''
\end{assumption}

Let $\ell: \cW \times \cZ \rightarrow \bR_+$ be a non-negative \emph{loss function}.
This function is a measure of how bad a hypothesis $W$ explains an instance $Z$.
Then, the \emph{population} and \emph{empirical risks} of a fixed hypothesis $w$ are defined as
\begin{align}
    L_{P_Z}(w) &\triangleq \bE_{P_Z}[\ell(w,Z)] \textnormal{ and}\\
    L_S(w) &\triangleq \frac{1}{N} \sum\nolimits_{Z \in S} \ell(w,Z),
\end{align}
respectively.
This way, the \emph{generalization error} of a learning algorithm $\bA$ is defined as the difference between the population and the empirical risks.
That is,
\begin{equation}
    \textnormal{gen}(W,S) \triangleq L_{P_Z}(W) - L_S(W).
\end{equation}

The following result bounds the absolute value of the \emph{expected} generalization error from above when the loss function is $\sigma$-sub-Gaussian.
We recall that a random variable $X$ is $\sigma$-sub-Gaussian if $\bE \big[ \exp \lambda (X - \bE[X]) \big] \leq \exp\left(\frac{\lambda^2 \sigma^2}{2}\right)$ for all $\lambda \in \bR$~\cite{buldygin1980sub}. 
Note that, by Hoeffding's 
{lemma}, if {$-\infty < L_{\textnormal{min}} \leq \ell(w,Z) \leq L_{\textnormal{max}} < \infty $ almost surely}, then $\ell(w,Z)$ is $(L_{\textnormal{max}} - L_{\textnormal{min}})/2$-sub-Gaussian.
Hence, if $|\cZ| < \infty$, then $\ell(w,Z)$ is $\sigma_w$-sub-Gaussian with 
$\sigma_w = \max_{a, b \in \cZ} |\ell(w,a)-\ell(w,b)|/2$.
This means that, if $\tilde{\sigma} = \sup_{w \in \mathcal{W}} \sigma_w$ exists, then $\ell(w,Z)$ is $\tilde{\sigma}$-sub-Gaussian for all $w \in \mathcal{W}$.

\begin{theorem}[\hspace{1sp}{\cite[Theorem 1]{xu2017information}}]
\label{th:xu_raginsky}
Let $S \in \cS$ be a dataset of $N$ instances $Z_n \in \cZ$ sampled i.i.d. from $P_Z$.
Let also $W \in \cW$ be a hypothesis obtained with an algorithm $\bA$, characterized by $P_{W|S}$.
Suppose $\ell(w,Z)$ is $\sigma$-sub-Gaussian under $P_Z$ for all $w \in \cW$.
Then,
\begin{equation}
    \big| \bE_{P_{W,S}}[\textnormal{gen}(W,S)] \big| \leq \sqrt{\frac{2 \sigma^2}{N} I(S;W)}
    \label{eq:xu_raginsky}
\end{equation}
\end{theorem}

Following Theorem~\ref{th:xu_raginsky}, we look for upper bounds on the mutual information $I(S;W)$, thus bounding the generalization error from above.
We require the mutual information upper bounds to grow slower than $N$, so that the resulting generalization error upper bounds tend to $0$ as $N$ increases.

\section{Preliminaries}
\label{sec:preliminaries}

In this section, we review some relevant theory needed to develop our findings. First, in Subsection~\ref{subsec:KL_div_MI}, we recall some results of the KL divergence and the mutual information. Then, in Subsection~\ref{subsec:method_types}, we introduce the method of types and its relationship with the concept of a dataset. Finally, in Subsection~\ref{subsec:gdp}, we present $\epsilon$-DP and $\mu$-GDP algorithms and their properties related to stability.

\subsection{Kullback--Leibler divergence and mutual information}
\label{subsec:KL_div_MI}

If we consider two random variables $X \in \mathcal{X}$ and $Y \in \mathcal{Y}$ with probability distributions $P_X$ and $P_Y$, and with conditional probability distribution $P_{Y|X}$, then
\begin{equation}
	I(X;Y) = \mathbb{E}_{P_X} \big[ \KL{P_{Y|X}}{P_Y} \big],
\end{equation}
which may also be denoted as $D_{\textnormal{KL}}(P_{Y|X}\, \|\, P_{Y}\, |\, P_X)$.
In order to bound the above quantity, we benefit from the following lemma, a consequence of the \emph{golden formula}~\cite[Theorem~3.3]{polyanskiy2014lecture}.

\begin{lemma}[{\hspace{1sp}\cite[Corollary 3.1]{polyanskiy2014lecture}}] 
\label{lemma:variational_bound}
Let $X \in \cX$ and $Y \in \cY$ be two random variables with {conditional probability distribution $P_{Y|X}$, and with marginal distributions $P_X$ and $P_Y$}, respectively. Then, 
\begin{equation}
    I(X;Y) \leq \mathbb{E}_{P_X} \big[ \KL{P_{Y|X}}{Q_Y} \big],
    \label{eq:variational_bound}
\end{equation}
where $Q_Y$ may be any probability distribution over $(\mathcal{Y}, \mathscr{Y})$. Moreover, \eqref{eq:variational_bound} is achieved with equality if and only if $Q_Y = P_Y$.
\end{lemma}

Note that for $X=S$ and $Y=W$, this Lemma states that, for any distribution $Q_W$ over $(\cW, \mathscr{W})$,
\begin{equation}
    I(S;W) \leq \bE_{P_S} \big[ \KL{P_{W|S}}{Q_W} \big],
    \label{eq:variational_bound_dataset}
\end{equation}
where the r.h.s. of~\eqref{eq:variational_bound_dataset} can equivalently be written as $\bE_{s \sim P_S} \big[ \KL{P_{W|S=s}}{Q_W} \big]$.
Hence, in this work we will focus on finding upper bounds on the relative entropy $\KL{P_{W|S=s}}{Q_W}$ for all $s \in \cS$ and some suitable distribution $Q_W$.

\begin{remark}
We focus on $\KL{P_{W|S=s}}{Q_W}$, instead of $I(S;W)$, since the former quantity appears in other expressions that characterize the generalization error, such as PAC-Bayesian bounds~\cite[Corollary 3]{hellstrom2020generalization}.
We comment further on this type of bounds in Section~\ref{subsec:applications}.
\end{remark}

\subsection{Method of types and datasets}
\label{subsec:method_types}

Let $Z^N = (Z_1, Z_2, \ldots, Z_N)$ be a sequence of i.i.d. random variables such that $Z_n$ has probability distribution $P_Z$ for all $n \in [N]$.
Then, $Z^N$ is also a random variable with probability distribution $P_{Z}^{\times N}$.

\begin{definition}[{\hspace{1sp}\cite[Section 11.1]{cover2012elements}}]
\label{def:type}
The \emph{type} $T_{Z^N}$, or \emph{empirical probability distribution}, of a sequence $Z^N \in \cZ^N$ is the relative proportion of occurrences of each symbol of $\cZ$.
That is, $T_{Z^N}(a) = N(a|Z^N) / N$ for all $a \in \cZ$, where $N(a|Z^N)$ is the number of times the symbol $a$ occurs in the sequence $Z^N$.
The set of all possible types of sequences of elements from $\cZ$ of length $N$ is denoted by $\cT_{\cZ,N}$.
\end{definition}

An interesting property of the types is that, although the number of elements in $\cZ^N$ scales exponentially in $N$, the total number of types only scales polynomically in $N$.
It is known that $|\cT_{\cZ,N}| \leq (N+1)^{|\cZ|}$ \cite[Theorem~11.1.1]{cover2012elements}.
This can be marginally improved by the following claim, which is tighter for finite $N$ and alphabets of small cardinality, especially binary alphabets.

\begin{claim}
\label{claim:num_types}
$|\cT_{\cZ,N}| \leq (N+1)^{|\cZ|-1}$, with equality if and only if $|\cZ|=2$.
\end{claim}
\begin{IEEEproof}
See Appendix~\ref{app:proof_claim_num_types}.
\end{IEEEproof}

Another interesting property of the type $T_{Z^N}$ of a sequence $Z^N$ is that it is equal to that of any permutation of the same sequence, i.e., $T_{Z^N} = T_{\mathcal{P}(Z^N)}$, where $\mathcal{P}$ is a random permutation.
Therefore, the type $T_{Z^N}$ uniquely identifies the dataset $S$ with elements $\lbrace Z_1, Z_2, \ldots, Z_N \rbrace$, where the order of the instances is irrelevant.
For this reason, we define the \emph{type of the dataset} $S$ as
\begin{equation}
	T_S \triangleq T_{\smash{\tilde{Z}^N}},
\end{equation}
where $\tilde{Z}^N$ is any sequence generated by all the $N$ instances of $S$.
In this way, we can define the distance between two datasets $S$ and $S'$ by means of their types $T_S$ and $T_{S'}$.

\begin{definition}
\label{def:dist}
 The \emph{distance} between two datasets $S$ and $S'$ is defined as the minimum number of instances that needs to be changed in $S$ to obtain $S'$.
 It is proportional to the total variation distance between the types $T_S$ and $T_{S'}$,
 \begin{equation*}
  d(S,S') \triangleq \frac{1}{2} \sum_{a\in\cZ} \big| N(a|S) - N(a|S') \big|
  = \frac{N}{2} \lnorm[\big]{T_S - T_{S'}}{1},
 \end{equation*}
 where $N(a|S)$ is the number of times the symbol $a$ occurs in the dataset $S$.
\end{definition}

\begin{remark}
The use of the method of types is justified according to Assumption~\ref{ass:algorithm} and the claim that there is a bijection between types and datasets, since a dataset $S$ is defined as a \emph{collection} of data samples, and not as a sequence.
Therefore, the order in which the samples are collected does not alter the dataset nor the distribution of the hypothesis that the algorithm outputs from the said dataset.
\end{remark}

\subsection{Privacy mechanisms}
\label{subsec:gdp}

A stochastic algorithm $\bA$ is said to be $\epsilon$-DP if, given a particular output of $\bA$, the maximum log-likelihood ratio between any two neighboring input datasets is upper-bounded by $\epsilon$.
Similarly, $\bA$ is said to be $\mu$-GDP if distinguishing between any two neighboring input datasets is, at least, as hard as distinguishing between two Gaussian random variables with unit variance and means at a distance $\mu$.
A proper definition of these frameworks is outside the scope of this work and we direct the interested reader to~\cite[Section 2]{dwork_dp_2014} ($\epsilon$-DP) and~\cite[Section 2]{dong2019gaussian} ($\mu$-GDP) for more details.

The concept of \emph{neighboring datasets} relates to a measure of distance between datasets. In this work, we consider two datasets to be neighbors, i.e., they have a distance $1$, if they differ in one element (see Definition~\ref{def:dist}).
Other works might consider this a distance $2$ since it involves removing one element from the first dataset and then adding a different element.
We want all datasets to have the same number of elements, and thus we believe that this second definition of distance is not appropriate.

Two important properties of $\epsilon$-DP and $\mu$-GDP algorithms are:
\begin{enumerate}
\item They are KL-stable\footnote{The definition of KL-stability in~\cite{bassily_algorithmic_2016} differs from~\eqref{eq:dp_distance_neighbor} and~\eqref{eq:gdp_distance_neighbor} in the explicit quantity on the r.h.s. of the bounds. However, the idea that the KL divergence is bounded remains.}~\cite[Definition 4.2]{bassily_algorithmic_2016}.
That is, given any two (fixed) neighboring datasets $s$ and $s'$, the KL divergence of their respective output distributions is bounded by~\cite[Lemma D.8]{dong2019gaussian}
 \begin{equation}
  \KL[\big]{\bA(s)}{\bA(s')} \leq \epsilon\tanh \left( \frac{\epsilon}{2} \right)
  \label{eq:dp_distance_neighbor}
 \end{equation}
if $\bA$ is $\epsilon$-DP, and by~\cite[Theorem.~2.10]{dong2019gaussian}
 \begin{equation}
  \KL[\big]{\bA(s)}{\bA(s')} \leq \frac{1}{2}\mu^2
  \label{eq:gdp_distance_neighbor}
 \end{equation}
if $\bA$ is $\mu$-GDP.

\item They possess group privacy.
Given datasets that are at a distance $k$, an $\epsilon$-DP algorithm $\bA$ is $k\epsilon$-DP~\cite[Theorem.~2.2]{dwork_dp_2014} and a $\mu$-GDP algorithm $\bA$ is $k\mu$-GDP~\cite[Theorem.~2.14]{dong2019gaussian}.
\end{enumerate}

A direct consequence of these two properties is condensed in the following claim.

\begin{claim}
\label{claim:diff_priv_distance}
Given an algorithm $\mathbb{A}$, if two (fixed) datasets $s$ and $s'$ are at a distance $k$, then the KL divergence of their respective output distributions is bounded from above as
\begin{equation}
	\KL[\big]{\bA(s)}{\bA(s')} \leq k\epsilon\tanh \left( \frac{k\epsilon}{2} \right) \leq k \epsilon
	\label{eq:dp_distance}
\end{equation}
if $\mathbb{A}$ is $\epsilon$-DP, and as
\begin{equation}
	\KL[\big]{\bA(s)}{\bA(s')} \leq \frac{1}{2} k^2 \mu^2
	\label{eq:gdp_distance}
\end{equation}
if $\mathbb{A}$ is $\mu$-GDP.
\end{claim}

\begin{remark}
\label{rm:dp_distance_tanh}
The tightness of the bound on the r.h.s. of~\eqref{eq:dp_distance} may be analyzed using the first term of the Taylor expansion of $\tanh(\cdot)$, which reveals that
\begin{equation}
 k\epsilon\tanh \left( \frac{k\epsilon}{2} \right) \leq \min \left\{ \frac{1}{2}k^2\epsilon^2, k\epsilon \right\}.
\end{equation}
We note that, if $\epsilon \geq 2/k$, the linear approximation in~\eqref{eq:dp_distance} is tighter than the quadratic one; moreover, this approximation becomes increasingly accurate as $k$ increases, given a fixed $\epsilon$.

Incidentally, the linear upper bound may also be obtained through the bound on the max-divergence in~\cite[Remark~3.1]{dwork_dp_2014}.
\end{remark}

\section{Main Results}
\label{sec:main_results}

In this section, we present our main results on generalization.
First, in Subsection~\ref{subsec:mixture_bound}, we introduce an auxiliary lemma that is essential to our results.
Then, in Subsection~\ref{subsec:simple_bound}, we show a generic upper bound for any learning algorithm.
Finally, in Subsection~\ref{subsec:gdp_bound}, we present upper bounds on private algorithms; more specifically, on $\epsilon$-DP and $\mu$-GDP algorithms.

\subsection{Auxiliary lemma}
\label{subsec:mixture_bound}

The following lemma, inspired by the variational approximation given in~\cite{hershey_2007_approx}, bounds the KL divergence between two probability distributions {$P$ and $Q$}, where the latter is a mixture probability distribution.

\begin{lemma}
\label{lemma:mixture_kl_ub}
Let $P$ and $Q$ be two probability distributions such that $P \ll Q$.
Let also $Q$ be a finite mixture of probability distributions such that $Q = \sum_b \omega_b Q_b$, where $\sum_b \omega_b = 1$, and $P \ll Q_b$ for all $b$.
Then, the following inequalities hold:
\begin{align}
	\KL{P}{Q} &\leq - \log \left( \sum\nolimits_b \omega_b \exp \big( {-} \KL{P}{Q_b} \big) \right) 
	\label{eq:lemma_ub_1} \\
	&\leq \min_b \big\lbrace \KL{P}{Q_b} - \log(\omega_b) \big\rbrace .
	\label{eq:lemma_ub_2} 
\end{align}
\end{lemma}
\begin{IEEEproof}
See Appendix~\ref{app:proof_lemma_mixture}.
\end{IEEEproof}

Given the element of the mixture $Q_b$, equation~\eqref{eq:lemma_ub_2} depicts the trade-off between its similarity with the distribution $P$ and its responsibility $\omega_b$.
Intuitively, the KL divergence between $P$ and $Q$ is bounded from above by the divergence between $P$ and the closest, most probable element of $Q$.
Particularly, if the weights $\omega_b$ are the same for every element of the mixture, then the bound~\eqref{eq:lemma_ub_2} is controlled by the element that is closest to $P$ in KL divergence.
This behavior of the bound will be useful in the proofs of the main results.
This bound is especially tight when one element of the mixture is either very probable, or much closer to $P$ than the others.
In the scenario where neither of these two conditions is met, the bound~\eqref{eq:lemma_ub_2} is loose and~\eqref{eq:lemma_ub_1} is preferred.

\subsection{A simple upper bound through the method of types}
\label{subsec:simple_bound}

The following proposition bounds from above the relative entropy $\KL{P_{W|S=s}}{Q_W}$ by means of Claim~\ref{claim:num_types} and
Lemma~\ref{lemma:mixture_kl_ub}.

\begin{proposition}
\label{prop:simple}
Let $S \in \mathcal{S}$ be a dataset of $N$ instances $Z_n \in \cZ$ sampled i.i.d. from {$P_Z$}.
Let also $W \in \mathcal{W}$ be a hypothesis obtained with an algorithm $\bA$, characterized by {$P_{W|S}$}.
Then, for all $s \in \cS$, there exists a distribution $Q_W$ over $(\cW,\mathscr{W})$ such that
\begin{equation}
 \KL{P_{W|S=s}}{Q_W} \leq (|\cZ|-1) \log(N+1).
 \label{eq:prop_simple_main}
\end{equation}
\end{proposition}
\begin{IEEEproof}
See Appendix~\ref{app:proof_prop_simple}.
\end{IEEEproof}

Note that in the proof of Proposition~\ref{prop:simple} we do not leverage the properties of Lemma~\ref{lemma:mixture_kl_ub} at their fullest, since we do not take advantage of the combination of the KL divergence and the mixture probabilities for the minimization.
However, if we note again that $I(S;W) \leq \bE_{s \sim P_S} \big[ \KL{P_{W|S=s}}{Q_W} \big]$, we observe how Lemma~\ref{lemma:mixture_kl_ub} allows us, naively, to obtain the same bound that one would obtain through the following decomposition of the mutual information:
\begin{align}
	I(S;W) &= h(S) - h(S|W)
	\stack{a}{\leq} h(S) \nonumber\\
	&\leq \log |\cS| 
	\stack{b}{\leq} (|\cZ|-1) \log(N+1), \label{eq:simple_obvious}
\end{align}
where $(a)$ is due to the non-negativity of the entropy for non-continuous random variables~\cite[Lemma 2.1.1]{cover2012elements}, and $(b)$ stems from Claim~\ref{claim:num_types} and the fact that there are as many datasets as possible types, i.e., $|\cS|=|\cT_{Z,N}|$.

\begin{remark}
\label{rem:multinomial}
We note that~\eqref{eq:simple_obvious} may be improved using the tighter bound on the entropy of a multinomial distribution from~\cite[Theorem~3.4]{kaji_bounds_2015}.
More precisely, assuming that $N$ is sufficiently large and after some  algebraic manipulations, the bound states that
\begin{equation}
    I(S;W) \preceq \frac{|\cZ| - 1}{2} \log \left( \frac{N}{|\cZ|} \right) + 2 |\cZ|, \label{eq:simple_obvious_multinomial}
\end{equation}
which grows more slowly than~\eqref{eq:simple_obvious} with respect to $N$, but has a more involved expression and interpretation.
\end{remark}

Both Proposition~\ref{prop:simple}, through Lemma~\ref{lemma:variational_bound}, and~\eqref{eq:simple_obvious} state the fact that, if $|\cZ|$ is finite, the number of possible datasets grows polynomially in $N$, which is independent of how the algorithm works.
In other words, even if the algorithm selects a different output hypothesis for each dataset, $I(S;W)$ cannot grow faster than logarithmically in $N$.
Therefore, according to Theorem~\ref{th:xu_raginsky}, if $\ell(w,Z)$ is $\sigma$-sub-Gaussian under {$P_Z$} for all $w \in \cW$, 
\begin{equation}
    \big| \bE_{P_{W,S}}[\textnormal{gen}(W,S)] \big| \leq \sqrt{2 \sigma^2 (|\cZ|-1) \frac{\log(N+1)}{N}},
	\label{eq:gen_error_ub_basic}
\end{equation}
which vanishes as $N$ grows, yielding Theorem~\ref{th:main_result_general}.
In the sequel, we restrict ourselves to algorithms that are $\epsilon$-DP or $\mu$-GDP, and thus we can improve upon this simple upper bound.

\subsection{Upper bounds on generalization for private algorithms}
\label{subsec:gdp_bound}

The following proposition bounds from above the relative entropy $\KL{P_{W|S=s}}{Q_W}$ of $\epsilon$-DP and $\mu$-GDP algorithms by means of Definition~\ref{def:type}, Lemma~\ref{lemma:mixture_kl_ub}, and Claim~\ref{claim:diff_priv_distance}.

\begin{proposition}
\label{prop:mi_mGDP}
Let $S \in \mathcal{S}$ be a dataset of $N$ instances $Z_n \in \cZ$ sampled i.i.d. from $P_Z$.
Let also $W \in \mathcal{W}$ be a hypothesis obtained with an algorithm $\bA$, characterized by $P_{W|S}$.
Then, for all $s \in \cS$, there exists a distribution $Q_W$ over $(\cW, \mathscr{W})$ such that:
\begin{enumerate}
\item If $\bA$ is $\epsilon$-DP and $\epsilon \leq 1$, then
\begin{equation}
	\KL{P_{W|S=s}}{Q_W} \leq (|\cZ|-1) \log \big(1 + e \epsilon N \big).
	\label{eq:prop_eDP_1_main}
\end{equation}
\item If $\bA$ is $\mu$-GDP and $\mu \leq 1/\sqrt{|\cZ|-1}$, then
\begin{multline}
	\KL{P_{W|S=s}}{Q_W} \leq \\ \frac{1}{2}(|\cZ|-1) \log \big( 1+ e (|\cZ| - 1) \mu^2 N^2 \big).
 	\label{eq:prop_mGDP_1_main}
\end{multline}
\end{enumerate} 
For lower privacy guarantees, i.e., $\epsilon > 1$ or $\mu > 1/\sqrt{|\cZ|-1}$, the upper bounds on $\KL{P_{W|S=s}}{Q_W}$ are no better than the one in Proposition~\ref{prop:simple}.
\end{proposition}
\begin{IEEEproof}
As previously mentioned, $\epsilon$-DP and $\mu$-GDP algorithms are smooth, or stable, in the sense that neighboring inputs produce similar outputs.
The proof of Proposition~\ref{prop:mi_mGDP} thus relies on a covering of the space of datasets and employs Lemma~\ref{lemma:mixture_kl_ub} to bound $\KL{P_{W|S=s}}{Q_W}$ using a particular mixture distribution $Q_W$.

Specifically, given a collection of datasets $\cS'$, the aforementioned distribution $Q_W$ is constructed as a uniform mixture of the probability distributions $\{P_{W|S=s}\}_{\smash{s\in\cS'}}$.
This collection is obtained by partitioning the space of datasets $\cS$ in equal-sized hypercubes and taking their ``central'' dataset.
In this way, the cover guarantees that the maximal distance inside a hypercube, i.e., $\max_{s \in \cS} \min_{s' \in \cS'} d(s,s')$, is bounded.
Then, we may bound the relative entropy $\KL{P_{W|S=s}}{Q_W}$ using Lemma~\ref{lemma:mixture_kl_ub} and Claim~\ref{claim:diff_priv_distance}.

See Appendix~\ref{app:proof_prop_mGDP_1} for details.
\end{IEEEproof}

We note that for a reasonably private algorithm, i.e., $\epsilon \leq 1/e$ or $\mu\leq 1/\sqrt{e(|\cZ|-1)}$, the upper bounds~\eqref{eq:prop_eDP_1_main} and~\eqref{eq:prop_mGDP_1_main} are tighter than the general one~\eqref{eq:prop_simple_main} (since $\sqrt{1+N^2} < 1+N$ for $N\geq 1$).

The upper bounds from Propositions~\ref{prop:simple} and~\ref{prop:mi_mGDP} can be tightened if a more accurate value for the number of hypercubes needed to cover the space of datasets is used.
However, this improvement in tightness comes at the expense of losing simplicity in the final expressions.
The following proposition summarizes the impact, in the previous results, of using a better approximation for the total number of covering hypercubes.

\begin{proposition}
\label{prop:mi_mGDP_2}
Let $S \in \mathcal{S}$ be a dataset of $N$ instances $Z_n \in \cZ$ sampled i.i.d. from $P_Z$.
Let also $W \in \mathcal{W}$ be a hypothesis obtained with an algorithm $\bA$, characterized by $P_{W|S}$.
Then, for all $s \in \cS$, there exists a distribution $Q_W$ over $(\cW, \mathscr{W})$ such that:
\begin{enumerate}
\item If $\bA$ is $\epsilon$-DP and $\epsilon \leq 1/N$, then
\begin{align}
    \MoveEqLeft[1]
	\KL{P_{W|S=s}}{Q_W} \nonumber \\
	&\leq (|\cZ|-1)(1+\epsilon N) - \frac{1}{2} \log 2 \pi (|\cZ|-1),
	\label{eq:prop_eDP_2_main_0}
\end{align}
and if $1/N < \epsilon \leq 1$, then
\begin{align}
    \MoveEqLeft[1]
	\KL{P_{W|S=s}}{Q_W} \nonumber \\
	&\leq (|\cZ|-1) \log \left(1 + \frac{2}{|\cZ|-1} \epsilon N  \right) \nonumber \\
	&\quad +(|\cZ|-1) \log \left( \frac{e^2}{2} \right) -\frac{1}{2} \log 2 \pi (|\cZ|-1).
	\label{eq:prop_eDP_2_main}
\end{align}
\item If $\bA$ is $\mu$-GDP and $\mu \leq 1/\big(N\sqrt{|\cZ|-1}\big)$, then
\begin{align}
    \MoveEqLeft[1]
    \KL{P_{W|S=s}}{Q_W} \nonumber \\
    &\leq (|\cZ|-1) \left( 1+ \frac{|\cZ|-1}{2} \mu^2 N^2 \right) \nonumber \\
    &\quad -\frac{1}{2} \log 2 \pi (|\cZ|-1),
    \label{eq:prop_mGDP_2_main_0}
\end{align}
and if $1/\big( N\sqrt{|\cZ|-1}\big) < \mu \leq 1/\sqrt{|\cZ|-1}$, then
\begin{align}
    \MoveEqLeft[1]
    \KL{P_{W|S=s}}{Q_W} \nonumber \\
    &\leq (|\cZ|-1) \log \left(1 + \frac{2}{\sqrt{|\cZ| -1}} \mu N \right) \nonumber \\
    &\quad +(|\cZ|-1) \log \left( \frac{ e^\frac{3}{2} }{ 2 } \right) -\frac{1}{2} \log 2 \pi (|\cZ|-1).
	\label{eq:prop_mGDP_2_main}
\end{align}
\item For any algorithm $\bA$, or if $\bA$ is $\epsilon$-DP and $\epsilon > 1$, or if $\bA$ is $\mu$-GDP and $\mu > 1/\sqrt{|\cZ|-1}$, then
\begin{align}
    \MoveEqLeft[1]
	\KL{P_{W|S=s}}{Q_W} \nonumber \\
	&\leq (|\cZ|-1) \log \left(1 + \frac{N}{|\cZ|-1} \right) \nonumber \\
	&\quad +(|\cZ|-1) -\frac{1}{2} \log 2 \pi (|\cZ|-1).
	\label{eq:prop_eDP_2_main_2}
\end{align}
\end{enumerate}
\end{proposition}
\begin{IEEEproof}
See Appendix~\ref{app:proof_prop_mGDP_2}.
\end{IEEEproof}

We note that, for large $\epsilon N$ or $\mu N$, the gap between~\eqref{eq:prop_eDP_1_main} and~\eqref{eq:prop_eDP_2_main}, and the gap between~\eqref{eq:prop_mGDP_1_main} and~\eqref{eq:prop_mGDP_2_main} grows as
\begin{equation}
	(|\cZ|-1) \log \left( \frac{ |\cZ|-1 }{ e }\right) +\frac{1}{2} \log 2 \pi (|\cZ|-1).
	\label{eq:diff_prop2_prop3}
\end{equation}
This highlights the benefit of Proposition~\ref{prop:mi_mGDP_2} for alphabets of large cardinality.

We also note that~\eqref{eq:prop_eDP_2_main_2} is a valid upper bound on $\KL{P_{W|S=s}}{Q_W}$ if the algorithms are not stable.
Similarly to Proposition~\ref{prop:simple}, the covering mixture used here takes every possible dataset.
However, this bound improves upon~\eqref{eq:prop_simple_main} by using a more exact value for the total number of datasets.
Consequently, the gap between~\eqref{eq:prop_simple_main} and~\eqref{eq:prop_eDP_2_main_2} also scales as~\eqref{eq:diff_prop2_prop3} for large $N$.

We further note that, according to Theorem~\ref{th:xu_raginsky}, if $\ell(w,Z)$ is $\sigma$-sub-Gaussian under $P_Z$ for all $w \in \cW$ and using the bounds~\eqref{eq:prop_eDP_1_main} and~\eqref{eq:prop_mGDP_1_main} from Proposition~\ref{prop:mi_mGDP} or the bounds~\eqref{eq:prop_eDP_2_main} and~\eqref{eq:prop_mGDP_2_main} from Proposition~\ref{prop:mi_mGDP_2}, the absolute value of the expected generalization error is asymptotically bounded as
\begin{equation}
  \big| \bE_{P_{W,S}}[\textnormal{gen}(W,S)] \big| \preceq \sqrt{2 \sigma^2 (|\mathcal{Z}|-1)\frac{\log(\gamma N)}{N}}
  \label{eq:gen_err_props_2_3}
\end{equation}
for both $\epsilon$-DP and $\mu$-GDP algorithms with $\epsilon = \gamma$ and $\mu = \gamma$, respectively.
Comparing the bounds~\eqref{eq:gen_error_ub_basic} and~\eqref{eq:gen_err_props_2_3}, we see the advantage of the private algorithms in reducing the generalization error, given that $\gamma<1$.

A final result, which is only valid for the first moment of $\KL{P_{W|S=s}}{P_W}$, i.e., $I(S; W) = \bE_{s \sim P_S}[\KL{P_{W|S=s}}{P_W}]$, is found in the following proposition, where the covering is not done on the whole space $\cS$ but rather the most likely datasets.

\begin{proposition}
\label{prop:mi_mGDP_3}
Let $S \in \mathcal{S}$ be a dataset of $N$ instances $Z_n \in \cZ$ sampled i.i.d. from $P_Z$.
Let also $W \in \mathcal{W}$ be a hypothesis obtained with an algorithm $\bA$, characterized by {$P_{W|S}$}.
\begin{enumerate}
\item If $\bA$ is $\epsilon$-DP and $\epsilon \leq 2$, then 
\begin{equation}
 I(S;W) \leq |\cZ| \log \big(1+ e \epsilon \sqrt{N \log N} \big) + 2|\cZ|\frac{\epsilon}{N},
 \label{eq:prop_eDP_3_main_1}
\end{equation}
while if $\epsilon>2$, 
\begin{equation}
 I(S;W) \leq |\cZ| \log \big( 1+ 2\sqrt{N\log N} \big) + 2|\cZ|\frac{\epsilon}{N}.
 \label{eq:prop_eDP_3_main_2}
\end{equation}
\item If $\bA$ is $\mu$-GDP and $\mu \leq 2/\sqrt{|\cZ|}$, then 
\begin{equation}
 I(S;W) \leq \frac{|\cZ|}{2} \log \left(1+ e |\cZ| \mu^2 N \log N\right) + |\cZ| \mu^2,
 \label{eq:prop_mGDP_3_main_1}
\end{equation}
while if $\mu>2/\sqrt{|\cZ|}$,
\begin{equation}
 I(S;W) \leq |\cZ| \log \big( 1+ 2\sqrt{N\log N} \big) + |\cZ| \mu^2.
 \label{eq:prop_mGDP_3_main_2}
\end{equation}
\end{enumerate}
\end{proposition}
\begin{IEEEproof}
See Appendix~\ref{app:proof_prop_mGDP_3}.
\end{IEEEproof}

{If we ignore the constant factors in~\eqref{eq:prop_eDP_3_main_1} and~\eqref{eq:prop_mGDP_3_main_1} and consider that $\ell(w,Z)$ is $\sigma$-sub-Gaussian under $P_Z$ for all $w \in \cW$, then the generalization error, according to Theorem~\ref{th:xu_raginsky}, is asymptotically bounded as}
\begin{equation}
  \big| \bE_{P_{W,S}}[\textnormal{gen}(W, S)] \big| \preceq \sqrt{2 \sigma^2 |\cZ| \frac{\log ( \gamma \sqrt{N \log N} )}{N}}
\end{equation}
for $\epsilon$-DP and $\mu$-GDP algorithms with $\epsilon = \gamma$ and $\mu = \gamma$, respectively.
If we compare this asymptotic behavior with the one from~\eqref{eq:gen_err_props_2_3} we see now how the privacy coefficients $\epsilon$ and $\mu$ multiply $\sqrt{N \log N}$ instead of $N$.
We note that $\sqrt{N \log N} < N$ for $N \geq 1$, which highlights the benefit of discriminating between typical and non-typical datasets.
This is our tightest result for private algorithms, and as such is summarized in Theorem~\ref{th:main_result_private}.

\begin{remark}
We note that the common Laplace or Gaussian mechanisms that assure $\epsilon$-DP and $\mu$-GDP add random noise of variance proportional to the inverse of the square of the parameter $\epsilon$ or $\mu$~\cite[Section~3.3]{dwork_dp_2014}, \cite[Theorem~2.7]{dong2019gaussian}.
If we compare the upper bounds obtained for $\epsilon$-DP and $\mu$-GDP algorithms in Propositions~\ref{prop:mi_mGDP}, \ref{prop:mi_mGDP_2}, and~\ref{prop:mi_mGDP_3}, we observe that their asymptotic behaviors appear to be almost identical by letting\footnote{In the case of Proposition~\ref{prop:mi_mGDP_3}, we would need $\mu = \epsilon /\sqrt{|\cZ|}$ to be more precise.} $\mu = \epsilon /\sqrt{|\cZ|-1}$.
This suggests that, even if both measures of privacy are not equivalent, $\mu$-GDP algorithms need to be ``noisier'' than $\epsilon$-DP algorithms, as $|\cZ|$ grows, in order to obtain similar generalization performance.

The main reason for this behavior is that our results depend on the stability of the algorithm, measured by the KL divergence between the hypotheses obtained by two datasets at a distance $k$ (see Claim~\ref{claim:diff_priv_distance}).
If we let $\epsilon = \mu = \gamma$, then the tighter upper bound of the KL divergence for $\gamma$-DP algorithms is always smaller than the upper bound for $\gamma$-GDP algorithms (see Remark~\ref{rm:dp_distance_tanh}).
Moreover, the looser upper bound of the KL divergence for $\gamma$-DP algorithms, i.e., $k\gamma$, is smaller than the upper bound for $\gamma$-GDP algorithms once $\gamma > 2/k$.
Therefore, the hypothesis distribution of $\mu$-GDP algorithms changes more rapidly with the distance between two datasets, thus making this kind of algorithms less smooth.
If we recall that the distance (in expectation) between two datasets grows with the size of the sample space $\cZ$, we can intuit the relationship between $\mu$ and $\epsilon$ in our results for generalization.
\end{remark}

\subsubsection{{A clarifying example}}
\label{subsubsec:clarifying_example}

{Imagine a medical setting where the doctors want to determine if an ICU patient should enter a treatment $B$ or should remain with the current treatment $A$.
To make their decision, they want the aid of a model $M$ 
that takes as input a set of hand-crafted features $X = (F_1, F_2, \dots, F_K)$ and predicts if the patient will survive a certain (fixed) time after entering treatment $B$.
An example of such a hand-crafted feature (which showcases its discrete nature) is the systolic blood pressure classified in the following intervals: lower than 120 mmMg, between 120 and 129 mmHg, between 130 and 139 mmHg, between 140 and 179 mmHg, or higher than 180 mmHg.}

{With this purpose, the doctors collect a set of $N$ historical records, $Z_n$, containing the aforementioned features, $X_n$, and if the patient survived, $Y_n$.
Then, they design the model $M$ (i.e., the hypothesis $W$) employing the historical records $\lbrace Z_1, \ldots, Z_N \rbrace$ (i.e., the dataset $S$) ensuring that it is $\epsilon$-DP to preserve the patients' anonymity, e.g., with private logistic regression~\cite[Section~4]{yu2014differentially}.
The model achieves a certain accuracy $\alpha \in [0,1]$.
Finally, they wonder how well, on average, the model will describe new patients' data for the loss function $\ell(m,z)=\mathbbm{1}(m(x)\neq y)$, where $\mathbbm{1}$ is the indicator function.}

\begin{figure}[h!]
    \centering
    \begin{subfigure}[b]{0.47\textwidth}
        \centering
        \includegraphics[width=\textwidth]{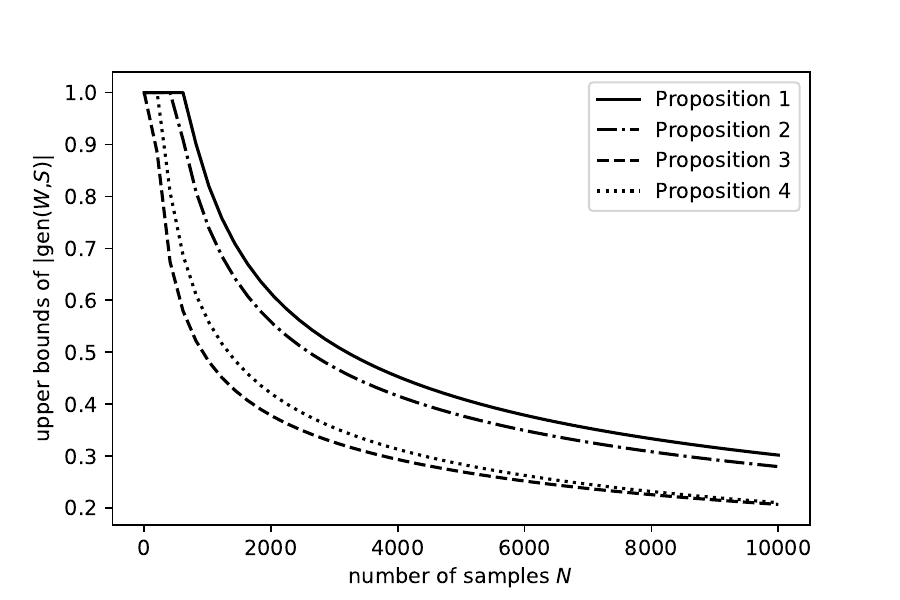}
    \end{subfigure}
    \begin{subfigure}[b]{0.47\textwidth}
        \centering
        \includegraphics[width=\textwidth]{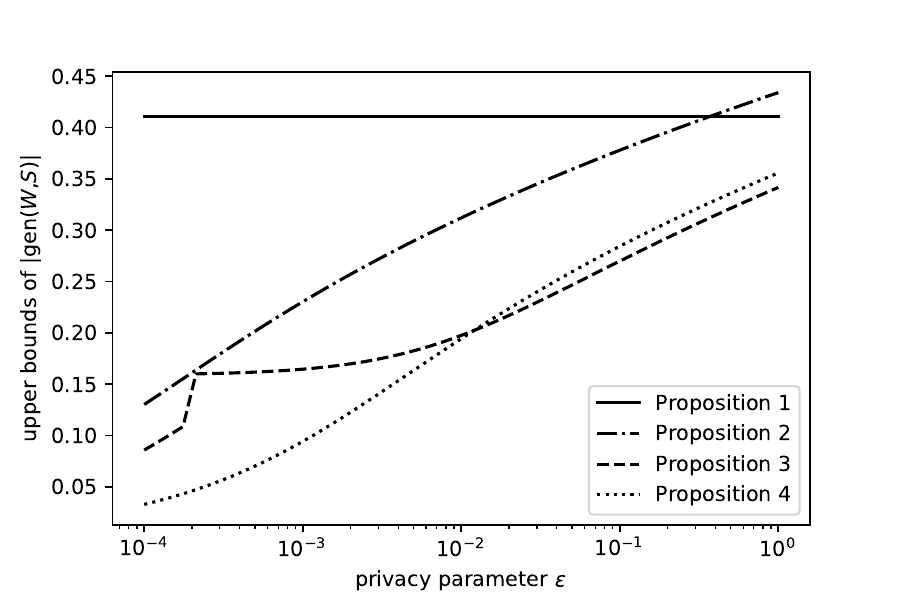}
    \end{subfigure}
    \caption{{Comparison of the different proposed bounds.}}
    \label{fig:comparison_within}
\end{figure}

{Figure~\ref{fig:comparison_within} shows the expected generalization error with the different presented propositions in two scenarios where the cardinality of the data, i.e., $|\cZ| = |\cX \times \cY|$, is fixed to $100$:
\begin{enumerate}[label=(\roman*)]
    \item in the first scenario, the privacy parameter $\epsilon$ is fixed to $0.1$ and the number of samples $N$ varies from $1$ to $10,000$; and
    \item in the second scenario, the number of samples $N$ is fixed to $5,000$ and the value of the privacy parameter $\epsilon$ varies from $10^{-4}$ to $1$.
\end{enumerate}
This figure showcases a scenario where non-vacuous generalization error bounds can be achieved, and depicts the trade-offs between privacy, number of samples, and generalization.
More specifically, the more private (smaller $\epsilon$) and the more samples we have, the better generalization guarantees.} 

\section{Discussion and Final Remarks}
\label{sec:discussion}

In this section, we review our results as well as discuss future directions these might open.
First, we contextualize our findings with some proposed applications.
Then, we compare our results with current upper bounds on private algorithms.
Finally, we conclude with the limitations of our work and present future steps to improve it.

\subsection{Applications}
\label{subsec:applications}

The results from Section~\ref{sec:main_results} may be used in different ways; we name a few in the sequel.

\subsubsection{{Bounding the relative entropy in other settings}}

{The bounds presented in Propositions~\ref{prop:mi_mGDP}, \ref{prop:mi_mGDP_2}, and~\ref{prop:mi_mGDP_3} focus on bounding the relative entropy between a conditional probability distribution $P_{Y|X}$ and the marginal distribution $P_Y$. Moreover, Lemma~\ref{lemma:mixture_kl_ub} bounds the relative entropy between two distributions $P$ and $Q$ when $Q$ is a finite mixture of probability distributions. Therefore, the presented results can be applied to any situation requiring to bound a relative entropy of these characteristics.}

{As an example, in order to obtain lower bounds on the minimax error, Fano's method applied to hypothesis testing requires to find upper bounds on the mutual information $I(S;V)$~\cite[Chapter~15]{wainwright2019high}. Here, $S = \lbrace Z_1, \ldots, Z_N \rbrace$ is still the available data and $V$ is a uniformly distributed random variable on a carefully designed set $\mathcal{V}$. It is common to employ the bound
\begin{align*}
    I(S;V) &= \frac{1}{|\cV|} \sum_{v \in \cV} \KL{P_{S|V=v}}{P_{S}} \\
    &\leq \frac{1}{|\cV|^2} \sum_{v,v' \in \cV} \KL{P_{S|V=v}}{P_{S|V=v'}}
\end{align*}
to characterize the mutual information. Note that in this case the distribution $P_S$ is precisely a mixture distribution $P_{S} = \frac{1}{|\mathcal{V}|} \sum_{v \in \cV} P_{S|V=v}$ and that the above expression is looser than the inequality resulting from~\eqref{eq:lemma_ub_1} due to Jensen's inequality. Therefore, the bounds presented in this work could be employed to better characterize the minimax risk in some situations. 
}

\subsubsection{Generalization error of other stable algorithms}

Although our main results deal with private algorithms, they can be adapted to other families of algorithms with stability or smoothness properties. 
That is, if the KL divergence between the distributions of the output hypotheses, produced by an algorithm fed with two datasets $s$ and $s'$ at a distance $k$, is bounded from above by some function $\phi(k)$, then Propositions~\ref{prop:mi_mGDP}, \ref{prop:mi_mGDP_2}, and~\ref{prop:mi_mGDP_3} can be replicated with this quantity in mind.
Note that when $k=1$, this property is known as $\sqrt{\phi(1)/2}$-KL-stability \cite[Definition 4.2]{bassily_algorithmic_2016}.

For example, the privacy framework of \emph{Rényi differential privacy}~\cite{mironov_renyi_2017} states that an algorithm $\bA$ is $(\alpha,\epsilon)$-RDP if $\RD{\bA(s)}{\bA(s')}\leq\epsilon$ for any two neighboring datasets $s$ and $s'$, where $\RD{P}{Q}$ is the Rényi $\alpha$-divergence of $P$ from $Q$.
Given the monotonicity of the Rényi $\alpha$-divergence with respect to $\alpha$, $\KL{\bA(s)}{\bA(s')} \leq \RD{\bA(s)}{\bA(s')}$ if $\alpha \geq 1$; the case $0< \alpha <1$ is more challenging but still possible to bound.
Furthermore, this framework can also be extended for group privacy, e.g., \cite[Proposition~2]{mironov_renyi_2017}, which enables the use of our results.

\subsubsection{Particularization to specific private algorithms}

The generalization error of specific learning algorithms that have been proved to be $\epsilon$-DP or $\mu$-GDP can be characterized.
For instance, the Noisy-SGD and Noisy-Adam~\cite[Algorithms~1 and~2]{bu2019deep}, private, regularized variants of the common stochastic gradient descent (SGD) and Adam~\cite[Algorithm~1]{kingma2014adam} algorithms, are approximately $\frac{B}{N} \sqrt{T (e^{1/{\nu}^2}-1)}$-GDP, where $B$ is the batch size, $T$ is the number of iterations made by the algorithm, and $\nu$ is the noise scale~\cite{bu2019deep}.
This example also highlights the benefit of private algorithms for generalization, given that the privacy parameter $\mu$, i.e., $\frac{B}{N} \sqrt{T (e^{1/{\nu}^2}-1)}$, decreases with the inverse of the number of samples $N$.

{Similarly, there are common algorithms like Markov chain Monte Carlo (MCMC) that can be proved to be $(\alpha,\epsilon)$-RDP~\cite{heikkila2019differentially} and others like support vector machines (SMVs) or classical logistic regression that can be adapted to be differentially private~\cite{yu2014differentially,rubinstein2009learning} with tunable privacy parameters.}

Employing the aforementioned characterization of the generalization capabilities of Rényi differentially private algorithms, one would be able to obtain bounds on the generalization error of MCMC.

\subsubsection{Other types of generalization bounds}

Other measures of the generalization error can also be characterized with the framework introduced in this work.
For example, the high-probability PAC-Bayesian bound from~\cite[Corollary 2]{hellstrom2020generalization}\footnote{{The bound from the corollary was corrected by the authors in https://gdurisi.github.io/files/2021/jsait-correction.pdf.}} states that, for all $\beta \in (0,1)$ and all distributions $Q_W$ over $(\cW, \mathscr{W})$ such that $P_{W,S} \ll Q_W \times P_S$, with probability no less than $1-\beta$ under $P_S$,
\begin{align}
    \MoveEqLeft[2]
    \big| \bE_{P_{W|S}}[\textnormal{gen}(W,S)] \big| \nonumber \\
    &\leq \sqrt{\frac{2 \sigma^2}{{N-1}} \left(\KL{P_{W|S}}{Q_W} + \log \frac{{\sqrt{N}}}{\beta} \right)}.
    \label{eq:disc_pac_bound}
\end{align}
We see that the bound~\eqref{eq:disc_pac_bound}, using the appropriate result from Section~\ref{sec:main_results} to characterize $\KL{P_{W|S=s}}{Q_W}$ for all $s \in \cS$, vanishes as $N$ increases.

\subsubsection{Randomized subsample setting}

Our ideas can be adapted to the randomized subsample setting introduced in~\cite{steinke2020reasoning}, which was originally conceived to bound the generalization error via a more robust information quantity than $I(W;S)$.
It is easy to see that the bound from Theorem~\ref{th:xu_raginsky} is vacuous in the case of continuous alphabets and a deterministic mapping between $S$ and $W$ since $P_{W,S} \,\not{\mkern -8mu\ll}\ P_W \times P_S$.

Here, a super-sample $\tilde{S} = \big( \tilde{Z}_1, \tilde{Z}_2, \ldots, \tilde{Z}_{2N}\big)$ of $2N$ i.i.d. samples from $P_Z$ is obtained first; then, the dataset $S$ is created by selecting between the samples $\tilde{Z}_i$ and $\tilde{Z}_{i+N}$ from $\tilde{S}$ based on the outcome of an independent Bernoulli random trial $U_i$; i.e., $Z_i = \tilde{Z}_{i + U_i N}$.
In this way, the \emph{empirical generalization error} may be defined as
\begin{align}
    \MoveEqLeft[2]
    \widehat{\textnormal{gen}}(W,\tilde{S},U) \triangleq 
    \frac{1}{N} \sum_{i=1}^N  \Big( \ell \big( W, \tilde{Z}_{i+(1-U_i)N} \big) - \ell \big( W,\tilde{Z}_{i+U_iN} \big)  \Big),
\end{align}
where $U = (U_1,U_2, \ldots, U_N)$ and $\bE_{P_{W,S}}[\textnormal{gen}(W,S)] = \bE_{P_{\smash{W,\tilde{S},U}}}[\widehat{\textnormal{gen}}(W,\tilde{S},U)]$.
The authors of~\cite{steinke2020reasoning} then obtain bounds on the generalization error which are based on $I(W; U| \tilde{S})$.
Unlike $I(W; S)$, this quantity is always finite; in particular, $I(W; U| \tilde{S}) \leq N\log 2$.

One possible way of using our framework in this setting is the following.
If we consider that the loss is bounded in $(a,b)$, we can look at the bounds from~\cite[Theorem~3.7]{haghifam2020sharpened} or \cite[Proposition~3]{rodriguez2020random} to obtain that
\begin{equation}
    \big| \bE_{P_{W,S}}[\textnormal{gen}(W,S)] \big| \leq \sqrt{-2(b-a)^2 \log\left( \frac{1}{2}+\frac{1}{2}e^{-\frac{\gamma^2}{2}} \right) },
    \label{eq:expected_generalization_cmi_bound_private}
\end{equation}
for any $\gamma$-DP or $\gamma$-GDP algorithm.

To see this, note that, for instance, \cite[Theorem~3.7]{haghifam2020sharpened} states that
\begin{align}
    \big| \bE_{P_{W,S}}&[\textnormal{gen}(W,S)] \big| \leq  \nonumber \\
    &\bE \Big[\sqrt{2(b-a)^2 \KL[\big]{P_{\smash{W|\tilde{S},U}}}{Q_{\smash{W|\tilde{S},U_{J^c}}}} } \Big],
\end{align}
where $J$ is a random index in $[N]$, $U_{J^c} = U \setminus U_J$, and $Q_{\smash{W|\tilde{S},U_{J^c}}}$ is any conditional distribution that maps realizations of $(\tilde{S},U_{J^c})$ to probability distributions on $(\cW, \mathscr{W)}$. Then, we may let $Q_{\smash{W|\tilde{S},U_{J^c}}} \triangleq (P_{W|S_0} + P_{W|S_1})/2$, where $S_0 \triangleq (S \setminus Z_J) \cup \tilde{Z}_J$ and $S_1 \triangleq (S \setminus Z_J) \cup \tilde{Z}_{J+N}$. Hence, since $P_{\smash{W|\tilde{S},U}} = P_{W|S}$ a.s. and either $S_0$ or $S_1$ will be equal to $S$ and the other will differ with $S$ in only one element, we may use Lemma~\ref{lemma:mixture_kl_ub} and Claim~\ref{claim:diff_priv_distance} to bound the relative entropy and obtain~\eqref{eq:expected_generalization_cmi_bound_private}.

\subsection{Comparison with current strict bounds}

The characterization of the generalization error for differentially private algorithms has predominantly been performed under the PAC-Bayesian framework. 
The current strict, i.e., not PAC-Bayesian, bounds on the expected generalization error for differentially private algorithms are derived either from (i) Theorem~\ref{th:xu_raginsky} using bounds on other measures of dependence between the hypothesis $W$ and the dataset $S$ or from (ii) the relationship between privacy and stability~\cite[Lemma 8]{JMLR:v17:15-313} and the bounds for stable algorithms~\cite{bousquet2002stability}.
We comment on a selection of such bounds in the sequel, and we collect them in Table~\ref{table:expected_gen_error} for reference.

\begin{table*}[t!]
    \centering
    \caption{Selection of recent bounds for the expected generalization error {(independent of the data cardinality)}}
    \begin{tabular}{lcc}
        \toprule
        & \multicolumn{2}{c}{Bounds according to loss} \\
        \cmidrule(r){2-3}
        Bounds based on & $\sigma$-sub-Gaussian & $(0,1)$-loss \\
        \midrule 
        Uniform stability~\cite[Lemma~8]{JMLR:v17:15-313} and~\cite{bousquet2002stability} & -- & $e^{\epsilon} -1$  \\
        Max-information~\cite[Thm.~7]{dwork_generalization_2015} & $\sigma \sqrt{2 \epsilon}$ & $\sqrt{\epsilon/2}$ \\
        Mutual information~\cite[Prop.~1.4, Thm.~1.10]{bun2016concentrated} & $\sigma \epsilon$ & $\epsilon/2$ \\
        \bottomrule
    \end{tabular}
    \label{table:expected_gen_error}
\end{table*}

Some of the said bounds may be obtained based on the \emph{max-information} $I_{\infty}(S;W)$ and $\beta$-\emph{approximate max-information} $I_{\infty}^\beta (S;W)$ between the training dataset $S$ of an algorithm and the hypothesis $W$ it produces~\cite[Definition~2]{dwork_dp_2014}.
In~\cite[Theorem~7]{dwork_generalization_2015} it is stated that an $\epsilon$-DP algorithm $\mathbb{A}$, characterized by {$P_{W|S}$}, trained on a dataset $S$ of $N$ i.i.d. samples, and that produced a hypothesis $W \in \mathcal{W}$ has a bounded max-information.
More precisely, $I_{\infty}(S;W) \leq \epsilon N$.
If we note that $I(S;W) \leq I_{\infty}(S;W)$ and we assume that $\ell(w,Z)$ is $\sigma$-sub-Gaussian under $P_Z$ for all $w \in \cW$, it is easy to see that the bounds on the max-information can be combined with Theorem~\ref{th:xu_raginsky} to find strict upper bounds on the generalization error.
The bounds leveraging $\beta$-approximate max-information cannot be compared with the proposed ones in this work, given that there is no strict relationship between this measure and the mutual information.
A stronger statement is proved in~\cite[Proposition 1.4 and Theorem 1.10]{bun2016concentrated}; that any $\epsilon$-DP algorithm $\bA$ has a bounded mutual information, $I(S;W) \leq \frac{1}{2} \epsilon^2 N$.

We conclude this part by comparing the aforementioned bounds (Table~\ref{table:expected_gen_error}) with those that can be obtained from Propositions~\ref{prop:mi_mGDP}, \ref{prop:mi_mGDP_2}, and~\ref{prop:mi_mGDP_3}, i.e., $\KL{P_{W|S=s}}{Q_W} \preceq (|\cZ|-1) \log(\epsilon N)$ and $\KL{P_{W|S=s}}{Q_W} \preceq |\cZ| \log(\epsilon \sqrt{N \log N})$, and the use of Theorem~\ref{th:xu_raginsky}.
We see that the bounds presented in this work are asymptotically tighter
since the expected generalization error vanishes as the number of samples grows (e.g., see Theorem~\ref{th:main_result_private}), as opposed to the bounds in Table~\ref{table:expected_gen_error}, for which the generalization error bound remains constant.
Nonetheless, our bounds depend on the cardinality of the input space, which is not the case for the other bounds;
those known bounds may thus provide a better characterization of the generalization error in the small sample regime. 

\subsubsection{{A numerical example}}

{In order to better clarify the comparison between the bounds in Table~\ref{table:expected_gen_error} and those presented in this work, we recover the example from Section~\ref{subsubsec:clarifying_example} and observe three different scenarios where three important parameters are varied:
\begin{enumerate}[label=(\roman*)]
    \item in the first scenario, the cardinality of the data $|\cZ|$ is set to $100$, the privacy parameter $\epsilon$ is set to $0.6$, and the number of samples $N$ varies from $1$ to $10,000$;
    \item in the second scenario, the cardinality of the data $|\cZ|$ is still fixed to $100$, but now the number of samples $N$ is fixed to $5,000$ and the privacy parameter $\epsilon$ varies from $10^{-4}$ to $2$; and
    \item in the third scenario, the privacy parameter $\epsilon$ is fixed to $0.6$, the number of data samples $N$ is also fixed to $5,000$, and the data cardinality $|\cZ|$ now varies from $2$ to $10,000$. 
\end{enumerate}}

\begin{figure}
    \centering
    \begin{subfigure}[b]{0.47\textwidth}
        \centering
        \includegraphics[width=\textwidth]{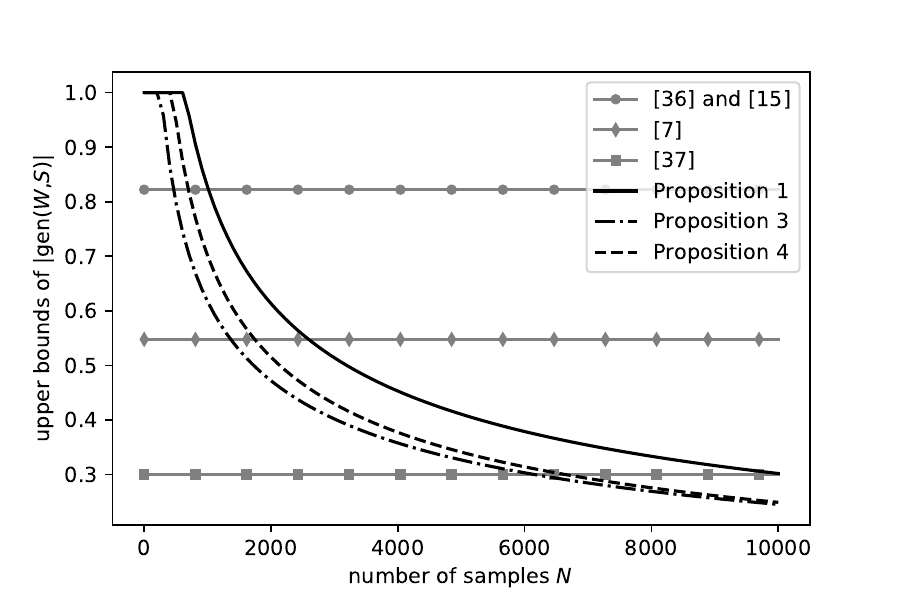}
    \end{subfigure}
    \begin{subfigure}[b]{0.47\textwidth}
        \centering
        \includegraphics[width=\textwidth]{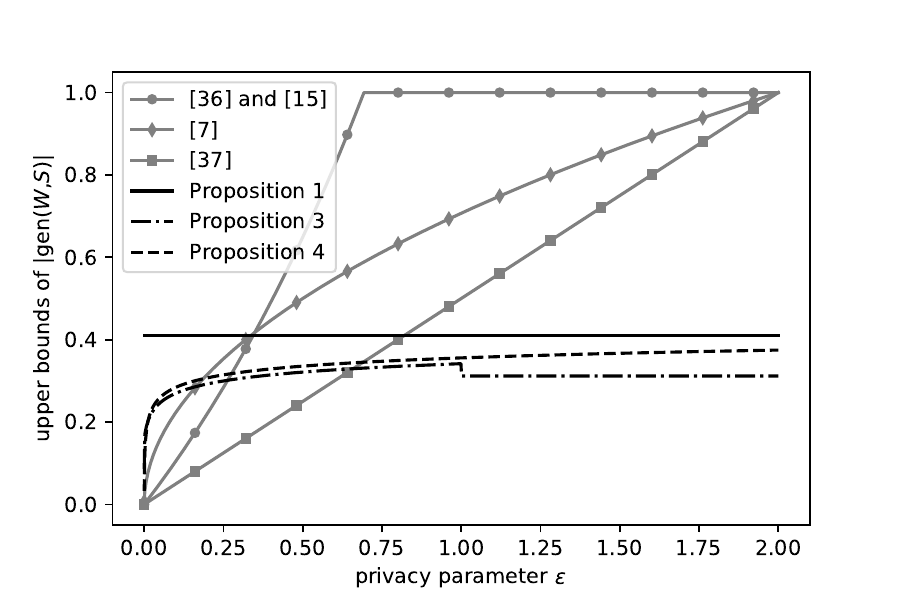}
    \end{subfigure}
    \begin{subfigure}[b]{0.47\textwidth}
        \centering
        \includegraphics[width=\textwidth]{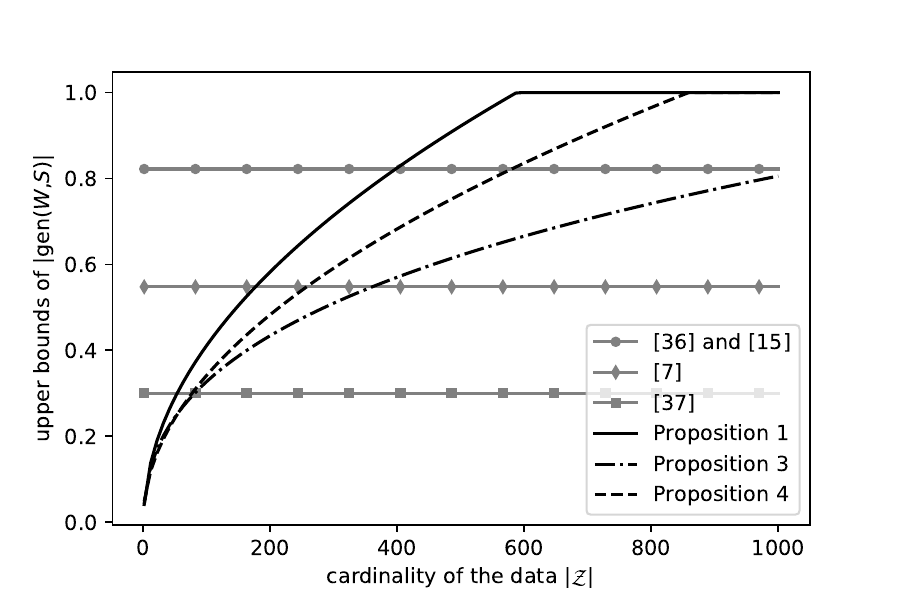}
    \end{subfigure}
    \caption{{Comparison of the proposed bounds with the bounds in Table I.}}
    \label{fig:comparison_between}
\end{figure}

{The results are presented in Figure~\ref{fig:comparison_between}.
In the first scenario, the results show that, as mentioned before, the bounds from Table~\ref{table:expected_gen_error} outperform the presented bounds in the small sample regime and are outperformed as the number of samples increases.
The results on the second scenario show how, for a moderate cardinality and number of samples, the presented bounds can outperform the bounds from Table~\ref{table:expected_gen_error} when the privacy parameter is not very small.
Nonetheless, for very private algorithms ($\epsilon \ll 1$) the bounds from Table~\ref{table:expected_gen_error} are generally tighter due to their linear dependency with $\epsilon$.
Finally, the results on the third scenario show how, for a moderate number of samples, the results from this work are only applicable in the small cardinality regime; when the cardinality of the data increases, other bounds like the ones in Table~\ref{table:expected_gen_error} are preferred.}

\subsection{Future work}

The results and ideas presented in this work build a framework for bounding the relative entropy between the conditional probability distribution of an algorithm's output hypothesis, given the dataset with which it was trained, and its marginal probability distribution.
Even though the obtained bounds have a good asymptotic behavior, they have an undesired offset produced by the cardinality of the sample space, $|\cZ|$.
A future step, then, is to use the ideas from Lemma~\ref{lemma:mixture_kl_ub} and Proposition~\ref{prop:mi_mGDP_3}, along with weak typicality, to extend this work to continuous random variables, thus producing tighter bounds on the generalization error.

Another future step is to adapt the approach of this paper to the randomized subsample setting from~\cite{steinke2020reasoning}, which was briefly described in Section~\ref{subsec:applications}.
In that setting, the generalization error can be bounded by a function of the conditional mutual information $I(W;U|\tilde{S})$, where $\tilde{S}$ is the super-sample and $U$ the vector of $N$ i.i.d. Bernoulli trials used to subsample $\tilde{S}$.
We may apply Lemma~\ref{lemma:mixture_kl_ub} to bound $I(W;U|\tilde{S})$ but, instead of covering the space of datasets as in our main results, we may cover the space of Bernoulli outcomes (also known as Hamming space), similar to~\cite[Section~V]{5992163}.
This type of bound would also be independent of the cardinality of $\cZ$.

\appendix

\subsection{Proof of Lemma~\ref{lemma:mixture_kl_ub}}
\label{app:proof_lemma_mixture}

We start the proof by defining the restricted measures $\tilde{Q}$ and $\tilde{Q}_b$ to be $Q$ and $Q_b$ in the support of $P$ and $0$ everywhere else. That is, 
\begin{equation}
    \tilde{Q} \triangleq \sum_b \omega_b \tilde{Q}_b = \sum_b \omega_b Q_b \chi_{\textnormal{supp}(P)},
\end{equation}
where $\chi_{A}$ is the characteristic function of a set $A$ (in this case, a collection of sets). We note that these definitions ensure that $\tilde{Q} \ll P$ and $\tilde{Q}_b \ll P$ for all $b$, while maintaining the property that $P \ll \tilde{Q}$ and $P \ll \tilde{Q}_b$ for all $b$, since $\textnormal{supp}(\tilde{Q}) = \textnormal{supp}(\tilde{Q}_b) = \textnormal{supp}(P)$ for all $b$. This way, we can manipulate the expression of the relative entropy as follows:
\begin{align}
    \KL{P}{Q} &= \bE_P \bigg[ \log \frac{dP}{dQ} \bigg] 
    \stack{a}{=} \bE_P \Bigg[ \log \frac{dP}{d\tilde{Q}} \Bigg] \nonumber \\ 
    &\stack{b}{=} - \bE_P \Bigg[ \log \frac{d\tilde{Q}}{dP} \Bigg] =- \bE_P \Bigg[ \log \frac{d \big( \sum_b \omega_b \tilde{Q}_b \big) }{dP} \Bigg] \nonumber \\
    &\stack{c}{=} - \bE_P \Bigg[ \log \Bigg( \sum_b \omega_b \frac{d \tilde{Q}_b}{dP} \Bigg) \Bigg], \nonumber
\end{align}
where (a) stems from the fact that the expectation will integrate $\frac{dP}{dQ}$ over the union of sets of the support of $P$, where $\frac{dP}{dQ} = \frac{dP}{d\tilde{Q}}$ $P$-a.e., and (b) and (c) stem from \cite[Exercise 9.27]{mcdonald1999course}.

Now, if we consider a set of positive coefficients $\lbrace \phi_b \rbrace_b$ such that $\sum_b \phi_b = 1$ we have that
\begin{align}
    \KL{P}{Q} &= - \bE_P \Bigg[ \log \Bigg( \sum_b \phi_b \frac{\omega_b}{\phi_b} \frac{d \tilde{Q}_b}{dP} \Bigg) \Bigg] \nonumber \\
    &\leq - \bE_P \Bigg[ \sum_b \phi_b \log \Bigg( \frac{\omega_b}{\phi_b} \frac{d \tilde{Q}_b}{dP} \Bigg) \Bigg] \nonumber \\ 
    &\triangleq \hat{D}_{\textnormal{KL}}(P\, \|\, \tilde{Q}\, ;\, \lbrace \phi_b \rbrace_b),
    \label{eq:approx_var_kl}
\end{align}
where the inequality stems from Jensen's inequality. 

We can now tighten the last inequality by minimizing the convex function $\hat{D}_{\textnormal{KL}}(P\, \|\, \tilde{Q}\, ;\, \lbrace \phi_b \rbrace_b)$ over the linear {constraint} $\sum_b \phi_b = 1$ with the Lagrangian $L(\lbrace \phi_b \rbrace_b, \lambda) = \hat{D}_{\textnormal{KL}}(P\, \|\, \tilde{Q}\, ;\, \lbrace \phi_b \rbrace_b) + \lambda(\sum_b \phi_b -1)$. The optimal value of the Lagrangian is given by,
\begin{equation}
    \frac{\partial L(\lbrace \phi_b \rbrace_b, \lambda)}{\partial \phi_b} = 0 \Leftrightarrow \phi_b^\star = \frac{\omega_b}{\exp(\lambda + 1)} \exp \big( {-} \KL{P}{\tilde{Q}_b} \big), \nonumber
\end{equation}
where we use the fact that $\frac{d\tilde{Q}_b}{dP} = \left(\frac{dP}{d\tilde{Q}_b}\right)^{-1}$ \cite[Exercise 9.27]{mcdonald1999course} and that $P \ll \tilde{Q}_b$ for all $b$.
Then, each $\phi_b^\star$ satisfies the linear {constraint} when
\begin{equation}
    \phi_b^\star = \frac{\omega_b \exp \big({-} \KL{P}{\tilde{Q}_b} \big)}{\sum_{b'} \omega_{b'} \exp \big({-} \KL{P}{\tilde{Q}_{b'}} \big)}. \nonumber
\end{equation}
Therefore, we can recover~\eqref{eq:lemma_ub_1} as follows:
\begin{align}
    \MoveEqLeft[2]
    \hat{D}_{\textnormal{KL}}(P\, \|\, \tilde{Q}\, ;\, \lbrace \phi_b^\star \rbrace_b) \nonumber \\
    &= -\bE_P \left[\sum_b \phi_b^\star \left( \log \left( \omega_b \frac{d \tilde{Q}_b}{dP} \right) - \log \phi_b^\star \right) \right]  \nonumber \\
    &= - \bE_P \left[\sum_b \phi_b^\star \left( \log \left( \omega_b \frac{d \tilde{Q}_b}{dP} \right) \right. \right. \nonumber \\
    & \qquad - \log \omega_b + \KL{P}{\tilde{Q}_b}  \nonumber \\
    & \qquad \left. \left. +  \log \left(\sum_{b'}\omega_{b'} \exp \big( {-} \KL{P}{\tilde{Q}_{b'}} \big) \right)  \right) \right]  \nonumber \\
    & = -\log \left(\sum_{b'}\omega_{b'} \exp \big( {-} \KL{P}{Q_{b'}} \big) \right), \label{eq:lemma_ub_1_proof}
\end{align}
where the last equality comes from the fact that $\sum_b \phi_b^\star = 1$, the fact that in $\KL{P}{\tilde{Q}_{b'}}$ the expectation w.r.t. $P$ will integrate $\log \frac{dP}{d\tilde{Q}_b}$ over the support of $P$, and the claim that
\begin{equation*}
    \bE_P \left[\sum_b \phi_b^\star \left( \log \left( \omega_b \frac{d \tilde{Q}_b}{dP} \right) - \log \omega_b + \KL{P}{\tilde{Q}_b} \right ) \right]
\end{equation*}
is 0, or equivalently, 
\begin{align*}
    \MoveEqLeft[2]
    \bE_P \left[\sum_b \phi_b^\star \left( \log \left( \omega_b \frac{d \tilde{Q}_b}{dP} \right) \right) \right] \\
    &= \sum_b \phi_b^\star \left(  \log \omega_b -\KL{P}{\tilde{Q}_b} \right),
\end{align*}
which stems again from the fact that $\frac{d\tilde{Q}_b}{dP} = \left(\frac{dP}{d\tilde{Q}_b}\right)^{-1}$ \cite[Exercise 9.27]{mcdonald1999course} and that $P \ll \tilde{Q}_b$ for all $b$. 
Finally, we can leverage the log-sum-exp bounds in~\eqref{eq:approx_var_kl} and~\eqref{eq:lemma_ub_1_proof} to obtain a more comprehensible upper bound on the relative entropy:
\begin{align}
    \MoveEqLeft[2]
	\KL{P}{Q} \nonumber \\ 
	&\leq  - \log \left( \sum_{b} \omega_{b} \exp \big( {-} \KL{P}{Q_{b}} \big) \right) \nonumber \\
	&= - \log \left( \sum_{b} \exp \big( {-} \KL{P}{Q_{b}} + \log (\omega_b) \big) \right) \nonumber \\
	&\leq - \log  \exp \max_b \big\lbrace {-} \KL{P}{Q_{b}} + \log (\omega_b) \big\rbrace \nonumber \\
	&\leq \min_b \big\lbrace \KL{P}{Q_b} - \log(\omega_b) \big \rbrace,
\end{align}
which is exactly the bound~\eqref{eq:lemma_ub_2}.
This concludes the proof of Lemma~\ref{lemma:mixture_kl_ub}.
\hfill\IEEEQED

\subsection{Proof of Proposition~\ref{prop:simple}}
\label{app:proof_prop_simple}

Let us consider the variational distribution of the hypothesis $Q_W$ to be a mixture of all the conditional distributions $P_{W|S}$ of the hypothesis given a dataset, with mixture probability (or responsibility) of $\omega_S$; that is, 
\begin{equation}
    Q_W = \sum_{s \in \cS} \omega_s P_{W|T_s},
\end{equation}
where we take into account that $P_{W|S=s} = P_{W|T_s}$ a.s. for all datasets $s \in \cS$, since a dataset $S$ with finite elements is completely characterized by its type $T_S$.

After that, we leverage Lemma~\ref{lemma:mixture_kl_ub} to obtain a more tractable upper bound. That is, 
\begin{align}
 \MoveEqLeft[2]
 \KL{P_{W|S=s}}{Q_W} \nonumber\\
 &\leq \min_{s' \in \cS} \left \lbrace \KL[\big]{ P_{W|S=s} }{ P_{ \smash{W|{T_{\smash{s'}}}} } } - \log \omega_{s'} \right \rbrace.
 \label{eq:prop1_bnd1}
\end{align}

A naive approach is to consider an equiprobable mixture $Q_W$, i.e., $\omega_s = |\cS|^{-1} = |\mathcal{T}_{\cZ,N}|^{-1}$ for all $s \in \cS$. Then, we have that
\begin{equation}
    \KL{P_{W|S=s}}{Q_W} \leq \log |\cS| \leq (|\cZ|-1) \log(N+1),
\end{equation}
where the second inequality stems from Claim~\ref{claim:num_types}.
This proves the proposition.
\hfill\IEEEQED

\subsection{Proof of Proposition~\ref{prop:mi_mGDP}}
\label{app:proof_prop_mGDP_1}

The proof takes advantage of the property that a private algorithm, e.g., $\epsilon$-DP or $\mu$-GDP, produces statistically similar outputs for neighboring datasets.
As in the proof of Proposition~\ref{prop:simple}, we employ Lemma~\ref{lemma:mixture_kl_ub} to obtain an upper bound on $\KL{P_{W|S=s}}{Q_W}$ but, this time, the mixture distribution $Q_W$ only uses some $P_{W|T_s}$.

We start by noting that all the types $T_s$ lie inside the unit hypercube $[0,1]^{|\cZ|-1}$. Although $T_s$ is a vector of $|\cZ|$ dimensions, since the last dimension is completely defined by the preceding $|\cZ|-1$, its possible values are located in a $(|\cZ|-1)$-dimensional subspace. This is the intuition behind Claim~\ref{claim:num_types}. 

To simplify the analysis, instead of studying the types, we focus on the vector of counts $N_s$, which is a scaled version of the type, i.e., $N_s(a) = N(a|s)$ for all $a \in \cZ$. As with the types, the first $|\cZ|-1$ dimensions completely define the vector of counts, and thus $N_s$ lies in a $[0,N]^{|\cZ|-1}$ hypercube.

\begin{figure}[t]
 \centering
 \includegraphics{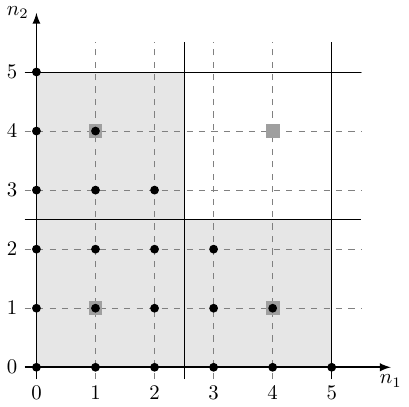}
 \caption{Example for $N=5$ samples, $|\cZ|=3$ dimensions, and $t=2$ parts per dimension.
 Here, $n_1$ and $n_2$ represent the counts of the first and second letters of the alphabet $\cZ$, i.e., $n_i = N_s(a_i)$.
 The black dots represent the possible datasets and the gray squares, the considered datasets for the mixture.
 The highlighted area represents the hypercubes that contain possible datasets.
 \label{fig:hypercubes_simplex_prop2}}
\end{figure}

We split the $[0,N]$ interval of each dimension in $1 \leq t\leq N$ parts, thus resulting in a $[0,N]^{|\cZ|-1}$ hypercube cover of $t^{|\cZ|-1}$ smaller hypercubes\footnote{We note that there are only $N+1$ coordinates in every dimension to choose as the center of a small hypercube. If we were to choose $t = N+1$, then we would have one hypercube per type, and thus we would not exploit the fact that the algorithm is smooth.}.
The types defined by the first $|\cZ|-1$ components of the vector of counts at the center of these smaller hypercubes are the ones selected to create the mixture (see Figure~\ref{fig:hypercubes_simplex_prop2} for an illustration).
We note that the side of every small hypercube has length 
\begin{equation}
	l \triangleq \frac{N}{t}
\end{equation}
and, more importantly, it has $l' \triangleq \lfloor l \rfloor + 1$ atoms.

If $l'$ is odd, we choose the central atom as the corresponding coordinate for the center of the small hypercube. On the other hand, if $l'$ is even, we enlarge the small hypercube in one unit and choose the center of this enlarged hypercube. Hence, the distance between the center and any other atom in the small hypercube, in only one of the dimensions is, at most,
\begin{equation}
	|N_s(a_i) - N_{s'}(a_i)| \leq \frac{\lfloor l \rfloor + 1}{2} \leq \frac{N}{2t} + \frac{1}{2}, \ \forall i \in \big[|\cZ|-1 \big],
\end{equation}
where $a_i$ is the $i$th element of $\cZ$. Therefore, for the first $|\cZ|-1$ components, the maximal distance between any vector of counts and the vector of counts at the center of the small hypercube is bounded by $(|\cZ|-1) (N / t + 1) /2$. In the worst case, all the counts in these first dimensions are off the center with the same sign (e.g., $N_s(a_i) - N_{s'}(a_i) = -(N/t + 1)/2$ for all $i \in \big[ |\cZ|-1 \big]$) and the last dimension (not directly included  but existent in the vector of counts) has to compensate for it. Hence, the distance on the last dimension of the vector of counts is bounded from above by
\begin{equation}
	|N_s(a_{|\cZ|}) - N_{s'}(a_{|\cZ|})| \leq \left(\frac{N}{2t} + \frac{1}{2}\right)(|\cZ|-1).
\end{equation}

The covering devised this way makes sure that the distance $d(s,s_i)$, as described in Definition~\ref{def:dist}, between any dataset $s$ belonging to the $i$th hypercube and its center $s_i$ is upper-bounded as follows,
\begin{equation}
 d(s,s_i) \leq \left(\frac{N}{2t} + \frac{1}{2}\right)(|\cZ|-1) \leq \frac{N}{t}(|\cZ| - 1),
 \label{eq:d_max}
\end{equation}
since $t \leq N$.
We then replicate the proof of Proposition~\ref{prop:simple} from~\eqref{eq:prop1_bnd1} using this new mixture and the properties of $\epsilon$-DP and $\mu$-GDP algorithms, i.e., Claim~\ref{claim:diff_priv_distance}.

\subsubsection{\texorpdfstring{$\epsilon$}{e}-DP algorithms}

We have that
\begin{align}
 \MoveEqLeft[2]
 \KL{P_{W|S=s}}{Q_W} \nonumber\\
 &\leq \min_{s' \in \cS'} \left \lbrace \KL[\big]{P_{W|S=s}}{ P_{W|T_{\smash{s'}}} } - \log \omega_{s'} \right \rbrace \nonumber\\
 &\leq \frac{N}{t} (|\cZ|-1) \epsilon + (|\cZ|-1) \log t,
 \label{eq:prop2_bnd1_dp}
\end{align}
where the second inequality follows from~\eqref{eq:dp_distance} in Claim~\ref{claim:diff_priv_distance}, \eqref{eq:d_max}, and the fact that there are at most $t^{|\cZ|-1}$ smaller hypercubes.
The value of $t$ that minimizes~\eqref{eq:prop2_bnd1_dp} is
\begin{equation}
 t = \epsilon N,
\end{equation}
and the following bound is obtained after replacing this value into~\eqref{eq:prop2_bnd1_dp},
\begin{equation}
	\KL{P_{W|S=s}}{Q_W} \leq (|\cZ|-1) \log \left( e \epsilon N \right).
 \label{eq:prop2_bnd2_dp}
\end{equation}
However, this result is only meaningful if the condition $1 \leq t \leq N$, as mentioned earlier in the proof, holds true, i.e., $1/N \leq \epsilon \leq 1$. 

On the one hand, if the optimal $t$ is such that $t > N$, we still need to choose $t=N+1$, as the maximum covering that can be designed selects one hypercube per type.
Proposition~\ref{prop:simple} has already addressed this situation and its result is tighter than~\eqref{eq:prop2_bnd1_dp}, with $t=N+1$, because of the loose upper bound~\eqref{eq:d_max}.
On the other hand, if the optimal $t$ is such that $t< 1$, we still need to choose $t=1$; a smaller $t$ implies that we are covering a volume larger than $[0,N]^{|\cZ|-1}$. The following upper bound is found in this way,
\begin{equation}
 \KL{P_{W|S=s}}{Q_W} \leq  (|\cZ| - 1) \epsilon N.
 \label{eq:prop2_bnd3_dp}
\end{equation}

We may further combine the bounds~\eqref{eq:prop2_bnd2_dp} and~\eqref{eq:prop2_bnd3_dp} into the more compact, albeit looser bound~\eqref{eq:prop_eDP_1_main}. We note that $x<\log(1+ex)$ if $x\leq 1$, which is equivalent to \eqref{eq:prop2_bnd3_dp}$<$\eqref{eq:prop_eDP_1_main} if $\epsilon N < 1$, precisely the region where~\eqref{eq:prop2_bnd3_dp} is valid.

\subsubsection{\texorpdfstring{$\mu$}{m}-GDP algorithms}

If we leverage~\eqref{eq:gdp_distance} from Claim~\ref{claim:diff_priv_distance} instead of~\eqref{eq:dp_distance} we now have that
\begin{align}
 \MoveEqLeft[2]
 \KL{P_{W|S=s}}{Q_W} \nonumber\\
 &\leq \min_{s' \in \cS'} \left \lbrace \KL[\big]{P_{W|S=s}}{ P_{W|T_{\smash{s'}}} } - \log \omega_{s'} \right \rbrace \nonumber\\
 &\leq \frac{1}{2} \frac{N^2}{t^2} (|\cZ|-1)^2 \mu^2 + (|\cZ|-1) \log t,
 \label{eq:prop2_bnd1_gdp}
\end{align}
where the value of $t$ that minimizes~\eqref{eq:prop2_bnd1_gdp} is now
\begin{equation}
 t = \sqrt{|\cZ|-1}\, \mu N,
\end{equation}
which yields
\begin{equation}
 \KL{P_{W|S=s}}{Q_W} \leq (|\cZ|-1) \log \left( \sqrt{ e (|\cZ| - 1)}\, \mu N \right).
 \label{eq:prop2_bnd2_gdp}
\end{equation}

If we operate analogously as with $\epsilon$-DP algorithms, we obtain that when $\mu \leq 1/\sqrt{|\cZ|-1}$, $\KL{P_{W|S=s}}{Q_W}$ is bounded from above by~\eqref{eq:prop_mGDP_1_main}, and that otherwise Proposition~\ref{prop:simple} is tighter.
This concludes the proof of Proposition~\ref{prop:mi_mGDP}.
\hfill\IEEEQED

\begin{remark}
A keen reader might notice that the upper bound on the KL divergence in~\eqref{eq:prop2_bnd1_dp} is not the tighter formula from~\eqref{eq:dp_distance} using $\tanh$.
Instead, we chose the linear upper bound for simplicity.

As mentioned in Remark~\ref{rm:dp_distance_tanh}, this approximation is better than a quadratic one if $\epsilon\geq 2/k$.
Had we chosen this other approximation, the result for $\epsilon$-DP would be the same as the one for $\mu$-GDP.
Comparing~\eqref{eq:prop2_bnd2_dp} and~\eqref{eq:prop2_bnd2_gdp}, with $\epsilon$ instead of $\mu$, we see that~\eqref{eq:prop2_bnd2_dp} is looser whenever $|\cZ|\leq 1+e$, i.e., only for binary and ternary alphabets.

For typical datasets, where $|\cZ|$ could be in the hundreds or more, the linear approximation is thus very tight.
Therefore, in the remaining proofs, we use the linear approximation in~\eqref{eq:dp_distance} for the KL divergence between outputs of $\epsilon$-DP algorithms.
\end{remark}

\subsection{Proof of Proposition~\ref{prop:mi_mGDP_2}}
\label{app:proof_prop_mGDP_2}

In the proof of Proposition~\ref{prop:mi_mGDP}, we devised a covering of the whole space $[0,N]^{|\cZ|-1}$ with $t^{|\cZ|-1}$ small hypercubes.
However, we observe that many of the hypercubes we designed contain no vector of counts. Particularly, there are only counts inside the hypervolume comprised between the origin and the $(|\cZ|-2)$-simplex; in the example from Figure~\ref{fig:hypercubes_simplex_prop2}, this volume is inside the highlighted area.
This simplex defines the manifold where the first $|\cZ|-1$ dimensions of the vector of counts sum up to $N$ in the $[0,N]^{|\cZ|-1}$ hypercube. Therefore, there are no possible vectors of counts above it since any dataset is restricted to $N$ samples; however, there are many vectors below it since the $|\cZ|$th dimension ``compensates'' for the unseen samples in the first $|\cZ|-1$ dimensions. 

For this reason, we only keep the hypercubes strictly needed to cover the hypervolume under the $(|\cZ|-2)$-simplex. This hypervolume is a hyperpyramid of height $N$ and its base is the hypervolume under the $(|\cZ|-3)$-simplex. Moreover, this hypervolume has $|\cZ|-1$ perpendicular edges of length $N$ which intersect at the origin.
For example, the hypervolume under the $1$-simplex is the right triangle with a vertex in the origin and two edges that go from $0$ to $N$ on both axes. The number of hypercubes needed to cover the hypervolume under the $(|\cZ|-2)$-simplex is given in the following lemma.

\begin{lemma}
\label{lemma:number_of_hypercubes_under_simplex}
The minimum number of hypercubes of a regular $t^{\times K}$ grid on the $[0,N]^K$ hypercube that covers the hypervolume under the $(K-1)$-simplex is
\begin{equation}
    S_K(t) = \frac{1}{K!} \frac{(t+K-1)!}{(t-1)!} \leq \frac{1}{K!} \left(t+\frac{K-1}{2}\right)^{K}.
    \label{eq:number_of_hypercubes_under_simplex}
\end{equation}
\end{lemma}
\begin{IEEEproof}
See Appendix~\ref{app:proof_lemma_hypercubes}.
\end{IEEEproof}

From Lemma~\ref{lemma:number_of_hypercubes_under_simplex} we know that we only need 
\begin{align}
    S_{|\cZ|-1}(t) &= \frac{1}{(|\cZ|-1)!} \frac{(t + |\cZ| - 2)!}{(t-1)!} \nonumber \\
    &\leq \frac{1}{(|\cZ|-1)!} \left(t + \frac{|\cZ|-2}{2} \right)^{|\cZ|-1}
    \label{eq:min_number_hypercubes}
\end{align}
hypercubes to cover all the possible vectors of counts in the $[0,N]^{|\cZ|-1}$ hypercube instead of the $t^{|\cZ|-1}$ used in Proposition~\ref{prop:mi_mGDP}.
We then replicate the proof of Proposition~\ref{prop:simple} from~\eqref{eq:prop1_bnd1} using the properties of $\epsilon$-DP and $\mu$-GDP algorithms, as we did in Proposition~\ref{prop:mi_mGDP}.

\subsubsection{\texorpdfstring{$\epsilon$}{e}-DP algorithms}

If we use the bound~\eqref{eq:dp_distance} in Claim~\ref{claim:diff_priv_distance}, \eqref{eq:d_max}, and the fact that the number of smaller hypercubes is bounded by~\eqref{eq:min_number_hypercubes}, we obtain that
\begin{align}
 \MoveEqLeft[2]
 \KL{P_{W|S=s}}{Q_W} \nonumber\\
 &\leq \min_{s' \in \cS'} \left \lbrace \KL[\big]{P_{W|S=s}}{ P_{W|T_{\smash{s'}}} } - \log \omega_{s'} \right \rbrace \nonumber\\
 &\leq \frac{N}{t} (|\cZ|-1) \epsilon + (|\cZ|-1) \log \left(t + \frac{|\cZ|-2}{2} \right) \nonumber \\
 &\quad - \log (|\cZ|-1)!.
 \label{eq:prop3_bnd1_dp}
\end{align}
The analytical expression for the value of $t$ that minimizes~\eqref{eq:prop3_bnd1_dp} is quite convoluted, so we study solutions of the form $t=\alpha \epsilon N$, $\alpha > 0$, based on the results from Proposition~\ref{prop:mi_mGDP}.
Then, in order to meet the condition $1 \leq t \leq N$, we need that $1/(\alpha N) \leq \epsilon \leq 1/\alpha$. 
Furthermore, we bound from above the last term in~\eqref{eq:prop3_bnd1_dp} using Stirling's formula, i.e., 
\begin{align}
-\log(|\cZ|-1)! &\leq -\frac{1}{2} \log 2 \pi (|\cZ|-1) \nonumber\\
    &\quad- (|\cZ|-1) \log(|\cZ|-1) + (|\cZ|-1).
    \label{eq:prop3_stirling}
\end{align}
This way, we obtain the upper bound 
\begin{align}
\MoveEqLeft[2]
    \KL{P_{W|S=s}}{Q_W} \nonumber\\
    &\leq (|\cZ|-1) \log \left( \frac{\alpha e^{\frac{1}{\alpha}+1}}{|\cZ|-1} \epsilon N + \frac{1}{2} \frac{|\cZ|-2}{|\cZ|-1} e^{\frac{1}{\alpha}+1} \right) \nonumber \\
    &\quad -\frac{1}{2} \log 2 \pi (|\cZ|-1) \nonumber \\ 
    &\leq (|\cZ|-1) \log \left( \frac{e^{\frac{1}{\alpha}+1}}{2} \left(1 + \frac{2\alpha}{|\cZ| -1} \epsilon N \right) \right) \nonumber \\
    &\quad -\frac{1}{2} \log 2 \pi (|\cZ|-1).
    \label{eq:prop3_alpha_bnd_dp}
\end{align}

Although it is possible to obtain an analytical formula for the optimal value of $\alpha$ in~\eqref{eq:prop3_alpha_bnd_dp}, we obtain no insights from it.
Instead, we provide a suboptimal value by analyzing the argument of the logarithm, in particular its derivative with respect to $\alpha$.
We observe that, as $\epsilon N/(|\cZ|-1)$ increases, the minimizing $\alpha$ tends to $1$ from above; if $\epsilon N= |\cZ|-1$, then $\alpha \approx 1.37$ is optimal, which is already quite close to the limit.
Therefore, we obtain~\eqref{eq:prop_eDP_2_main} by setting $\alpha$ to 1.

Having set $\alpha=1$, if the optimal $t\ (=\epsilon N)$ is such that $t < 1$, we fix $t = 1$ in~\eqref{eq:prop3_bnd1_dp} as explained in the proof of Proposition~\ref{prop:mi_mGDP}.
After some algebraic manipulations, we obtain the bound~\eqref{eq:prop_eDP_2_main_0}.

\subsubsection{\texorpdfstring{$\mu$}{m}-GDP algorithms}

If we leverage~\eqref{eq:gdp_distance} in Claim~\ref{claim:diff_priv_distance} instead of~\eqref{eq:dp_distance}, we have that
\begin{align}
 \MoveEqLeft[2]
 \KL{P_{W|S=s}}{Q_W} \nonumber\\
 &\leq  \min_{s' \in \cS'} \left \lbrace \KL[\big]{P_{W|S=s}}{ P_{W|T_{\smash{s'}}} } - \log \omega_{s'} \right \rbrace \nonumber\\
 &\leq \frac{1}{2} \frac{N^2}{t^2} (|\cZ|-1)^2 \mu^2 + (|\cZ|-1) \log \left(t + \frac{|\cZ|-2}{2} \right) \nonumber \\
 &\quad - \log (|\cZ|-1)!.
 \label{eq:prop3_bnd1_gdp}
\end{align}
As for $\epsilon$-DP algorithms, the analytical expression for the value of $t$ that minimizes~\eqref{eq:prop3_bnd1_gdp} is quite convoluted, so we study solutions of the form  $t=\alpha \sqrt{|\cZ|-1}\, \mu N$, based on the results from Proposition~\ref{prop:mi_mGDP}.
Then, in order to meet the condition $1 \leq t \leq N$, we need that
\begin{equation}
    \frac{1}{\alpha N \sqrt{|\cZ|-1}} \leq \mu \leq \frac{1}{\alpha \sqrt{|\cZ|-1}},
\end{equation}
and we obtain the upper bound
\begin{align}
\MoveEqLeft[2]
    \KL{P_{W|S=s}}{Q_W} \nonumber\\
    &\leq (|\cZ|-1) \log \left( \frac{\alpha e^{\frac{1}{2\alpha^2}+1}}{\sqrt{|\cZ|-1}} \mu N + \frac{1}{2} \frac{|\cZ|-2}{|\cZ|-1} e^{\frac{1}{2\alpha^2}+1} \right) \nonumber \\
    &\quad -\frac{1}{2} \log 2 \pi (|\cZ|-1) \nonumber \\ 
    &\leq (|\cZ|-1) \log \left( \frac{e^{\frac{1}{2\alpha^2}+1}}{2} \left( 1 + \frac{2\alpha}{\sqrt{|\cZ|-1}} \mu N \right) \right) \nonumber \\
    &\quad-\frac{1}{2} \log 2 \pi (|\cZ|-1).
    \label{eq:prop3_alpha_bnd_gdp}
\end{align}

We analyze again the derivative with respect to $\alpha$ of the argument of the logarithm in~\eqref{eq:prop3_alpha_bnd_gdp}. We now observe that, as $\mu N / \sqrt{|\cZ|-1}$ increases, the minimizing $\alpha$ also tends to 1 from above; if $\mu N = \sqrt{|\cZ| -1}$, $\alpha \approx 1.19$ is optimal, which is already quite close to the limit. Therefore, we obtain~\eqref{eq:prop_mGDP_2_main} by setting $\alpha$ to 1.

If the optimal value of $t$ is such that $t < 1$, we proceed similarly as with $\epsilon$-DP algorithms and we obtain the remaining bound.

\subsubsection{General algorithms}

For both $\epsilon$-DP and $\mu$-GDP, if the privacy guarantees are not good enough, the minimizations in~\eqref{eq:prop3_bnd1_dp} and~\eqref{eq:prop3_bnd1_gdp} result in an optimal value of $t> N$.
As explained in the proof of Proposition~\ref{prop:mi_mGDP}, we choose $t=N+1$ as, otherwise, we would have more smaller hypercubes than types.
In this case, there is one type per hypercube and we may replicate the proof of Proposition~\ref{prop:simple} and continue from~\eqref{eq:prop1_bnd1}.
That is,
\begin{align}
 \MoveEqLeft[2]
 \KL{P_{W|S=s}}{Q_W} \nonumber\\
 &\leq \min_{s' \in \cS} \left \lbrace \KL[\big]{P_{W|S=s}}{ P_{W|T_{\smash{s'}}} } - \log \omega_{s'} \right \rbrace \nonumber\\[.25em]
 &\stack{a}{=} \log S_{|\cZ|-1}(N+1) \nonumber\\
 &\leq (|\cZ|-1) \log \left(N+1 + \frac{|\cZ|-2}{2} \right) - \log (|\cZ|-1)!,
\end{align}
where $(a)$ is due to $\omega_s = |\cS|^{-1} = \big(S_{|\cZ|-1}(N+1)\big)^{-1}$ for all $s \in \cS$. 
After replacing the factorial with Stirling's approximation~\eqref{eq:prop3_stirling} and performing some algebraic manipulations, we obtain the bound~\eqref{eq:prop_eDP_2_main_2}.
This concludes the proof of Proposition~\ref{prop:mi_mGDP_2}.
\hfill\IEEEQED

\subsection{Proof of Proposition~\ref{prop:mi_mGDP_3}}
\label{app:proof_prop_mGDP_3}

In the proofs of Propositions~\ref{prop:mi_mGDP} and~\ref{prop:mi_mGDP_2}, we designed a covering of the whole space of datasets to approximate the marginal distribution of the output hypothesis.
However, as $N$ increases, the datasets that are more likely to be chosen accumulate on a certain region of space.
Similarly to typical sequences, we may define typical datasets given the connection between datasets and types.
We recall the definition of the \emph{strong typical} set:
\begin{align}
 \cT_\varepsilon^N(Z) &\triangleq \big\lbrace Z^N: \big| T_{Z^N}(a) - P_Z(a) \big| \leq \varepsilon \textnormal{ if } P_Z(a)>0, \nonumber \\
 &\qquad N(a|Z^N) =0\textnormal{ otherwise} \big\rbrace,
 \label{eq:prop4_typical_set}
\end{align}
where $T_{Z^N}(a)$ and $N(a|Z^N)$ are defined in Definition~\ref{def:type}.
For simplicity, we denote the strong typical set as $\cT$ in the sequel.
We also assume that $P_Z(a)>0$ for all $a \in \cZ$; otherwise, we may eliminate the elements with zero probability and reduce the cardinality of $\cZ$.

Directly from the definition in~\eqref{eq:prop4_typical_set}, we have that, for all $Z^N \in \cT$,
\begin{equation}
 \big| N(a|Z^N) -N P_Z(a) \big| \leq N \varepsilon,\ \forall a\in\cZ,
\end{equation}
and by Hoeffding's inequality and the union bound,
\begin{equation}
  P_{Z}^{\times N}( \cZ^{N} \setminus \cT) \leq 2|\cZ| \exp(-2N\varepsilon^2).
\end{equation}
If we choose $\varepsilon=\sqrt{\frac{\log N}{N}}$, we get that the typical vector of counts $N_S$ is found inside a $|\cZ|$-dimensional hypercube\footnote{Here we note that the covering is performed in a $|\cZ|$-dimensional space, unlike the previous proofs where we covered $(|\cZ|-1)$-dimensional spaces. Although this leads to a looser bound, the final expression is more manageable.}
 of side
\begin{equation}
 l_{\cT} \leq 2\sqrt{N \log N},
 \label{eq:prop4_l_T}
\end{equation}
with a probability
\begin{equation}
 P_S(\cT) \geq 1 -2|\cZ| N^{-2}.
 \label{eq:prop4_prob_s}
\end{equation}

We may now devise a covering of the set $\cT$ by splitting each dimension in $t$ parts. Thus, the side of each small hypercube has length
\begin{equation}
 l \triangleq \frac{l_{\cT}}{t} \leq \frac{2\sqrt{N \log N}}{t},
\end{equation}
where $t\leq2\sqrt{N \log N}$ or, otherwise, we have less than one type per hypercube.
Following a similar analysis as in the previous proofs, we have that, for every dimension $i \in \big[|\cZ|\big]$, the distance between the center and any other atom in the small hypercube is bounded as follows,
\begin{align}
|N_s(a_i) - N_{s'}(a_i)| &\leq \frac{\lfloor l \rfloor + 1}{2} \nonumber\\
&\leq \frac{\sqrt{N \log N}}{t} + \frac{1}{2} \nonumber\\
&\leq \frac{2\sqrt{N \log N}}{t}.
\end{align}
where the last inequality is due to the bound on $t$.
Therefore, the distance $d(s,s_i)$, as described in Definition~\ref{def:dist}, between any dataset $s$ belonging to the $i$th hypercube and its center $s_i$ is bounded as
\begin{equation}
 d(s,s_i) \leq \frac{\sqrt{N \log N}}{t} |\cZ|.
 \label{eq:prop4_d_max}
\end{equation}

We now proceed to bound the desired mutual information using 
Lemma~\ref{lemma:variational_bound} and the law of total expectation,
\begin{align}
I(S;W) &\leq \bE_{s \sim P_S} \big[ \KL{P_{W|S=s}}{Q_W} \mid S \in \cT \big] P_S(\cT) \nonumber\\
 & + \bE_{s \sim P_S} \big[ \KL{P_{W|S=s}}{Q_W} \mid S \notin \cT \big] P_S(\cS \setminus \cT),
\label{eq:prop4_total_exp}
\end{align}
where the second term on the r.h.s of~\eqref{eq:prop4_total_exp} may be bounded from above using~\eqref{eq:prop4_prob_s} and~\eqref{eq:lemma_ub_1} from Lemma~\ref{lemma:mixture_kl_ub}.
More precisely, for all $s \notin \cT$,
\begin{align}
\MoveEqLeft
\KL{P_{W|S=s}}{Q_W} \nonumber \\
	&\stack{a}{\leq} - \log \left( \sum\nolimits_{s'} \omega_{s'} \exp \big( {-} \KL{P_{W|S=s}}{P_{W|S=s'}} \big) \right) \nonumber \\
	&\stack{b}{\leq} \sum\nolimits_{s'} \omega_{s'} \KL{P_{W|S=s}}{P_{W|S=s'}} \nonumber \\ 
	&\leq \max_{s'} \big\lbrace \KL{P_{W|S=s}}{P_{W|S=s'}} \big\rbrace,
	\label{eq:worst_case_div}
\end{align}
where $(a)$ is due to~\eqref{eq:lemma_ub_1} and considering the mixture $Q_W$ as the weighted sum of $P_{W|S=s'}$, where each $s'$ is the dataset at the center of the covering hypercubes; and $(b)$ stems from Jensen's inequality.
Next, we may leverage Claim~\ref{claim:diff_priv_distance} in order to find the upper bounds for $\epsilon$-DP and $\mu$-GDP algorithms.

\subsubsection{\texorpdfstring{$\epsilon$}{e}-DP algorithms}

If we assume the worst case in bound~\eqref{eq:dp_distance} from Claim~\ref{claim:diff_priv_distance}, we have that~\eqref{eq:worst_case_div} leads to
\begin{equation}
	\KL{P_{W|S=s}}{Q_W} \leq \epsilon N.
	\label{eq:worst_case_div_dp}
\end{equation}
Therefore, since~\eqref{eq:worst_case_div} does not depend on $S$, and taking into account~\eqref{eq:prop4_prob_s}, we have that
\begin{equation}
	\bE_{s \sim P_S} \big[ \KL{P_{W|S=s}}{Q_W} \mid S \notin \cT \big] P_S(\cS \setminus \cT) \leq 2 |\cZ| \frac{\epsilon}{N}.
	\label{eq:second_term_rhs_prop4_dp}
\end{equation}

The first term on the r.h.s. of~\eqref{eq:prop4_total_exp} may be analyzed similarly as in the previous proofs but considering only the covering of $\cT$; this reduced covering determines the maximum distance~\eqref{eq:prop4_d_max} inside each small hypercube.
In other words,
\begin{align}
\MoveEqLeft
\bE_{s \sim P_S} \big[ \KL{P_{W|S=s}}{Q_W} \mid S \in \cT \big] \nonumber\\
 &\leq \frac{\sqrt{N \log N}}{t} |\cZ| \epsilon +|\cZ| \log t.
 \label{eq:prop4_partial_exp}
\end{align}
The value of $t$ that minimizes this expression is 
\begin{equation}
t = \sqrt{N \log N} \epsilon,
\label{eq:prop4_opt_t_dp}
\end{equation}
where we need to verify that $1 \leq t \leq 2 \sqrt{N \log N}$. If this condition holds, we may replace~\eqref{eq:prop4_opt_t_dp} in~\eqref{eq:prop4_partial_exp} and, jointly with~\eqref{eq:prop4_total_exp} and~\eqref{eq:second_term_rhs_prop4_dp}, obtain
\begin{equation}
I(S;W) \leq |\cZ| \log(e \epsilon \sqrt{N \log N}) + 2|\cZ| \frac{\epsilon}{N}.
\label{eq:prop4_mut_inf_1_dp}
\end{equation}
Given the constraint on $t$, the range of $\epsilon$ for which this bound is valid is 
\begin{equation}
\frac{1}{\sqrt{N \log N}} \leq \epsilon \leq 2.
\end{equation}

If $\epsilon < 1 /\sqrt{N \log N}$, which corresponds to an optimal $t$ such that $t < 1$ according to~\eqref{eq:prop4_opt_t_dp}, we choose $t=1$ as it was noted in the proof of Proposition~\ref{prop:mi_mGDP}. In this case, the mutual information is bounded as follows, 
\begin{equation}
	I(S;W) \leq |\cZ| \epsilon \sqrt{N \log N} + 2|\cZ| \frac{\epsilon}{N}.
	\label{eq:prop4_mut_inf_2_dp}
\end{equation}
We then proceed to combine~\eqref{eq:prop4_mut_inf_1_dp} and~\eqref{eq:prop4_mut_inf_2_dp} into a single, more compact bound. Using a similar argument as in the proof of Proposition~\ref{prop:mi_mGDP}, we obtain~\eqref{eq:prop_eDP_3_main_1}.

Finally, if $\epsilon > 2$, which corresponds to $t > 2 \sqrt{N \log N}$, we fix $t = 2 \sqrt{N \log N} + 1$. In this case, we are including each dataset in $\mathcal{T}$ into the mixture. Instead of replacing this value of $t$ in~\eqref{eq:prop4_partial_exp}, we note that a uniform covering of $\cT$ results in
\begin{align}
\MoveEqLeft
 \bE_{s \sim P_S} \big[ \KL{P_{W|S=s}}{Q_W} \mid S \in \cT \big] P_S(\cT) \nonumber\\
 &\stack{a}{\leq} \sum_{s \in \cT} P_S(s) \min_{s' \in \cT} \left \lbrace \KL[\big]{P_{W|S=s}}{P_{W|{T_{s'}}}} - \log \omega_{s'} \right \rbrace \nonumber \\
 &= P_S(\cT) \log |\cT| \nonumber\\
 &\stack{b}{\leq} |\cZ| \log \big( 1+ 2\sqrt{N\log N} \big),
 \label{eq:prop4_large_epsilon}
\end{align}
where $(a)$ is due to Lemma~\ref{lemma:mixture_kl_ub}; and $(b)$ follows from the fact that $\cT$ is contained by a $\cZ$-dimensional hypercube of side $2\sqrt{N\log N}$ as stated in~\eqref{eq:prop4_l_T}. Combining~\eqref{eq:prop4_total_exp}, \eqref{eq:second_term_rhs_prop4_dp}, and~\eqref{eq:prop4_large_epsilon}, we obtain~\eqref{eq:prop_eDP_3_main_2}.

\subsubsection{\texorpdfstring{$\mu$}{m}-GDP algorithms}

If we assume now the worst case in bound~\eqref{eq:gdp_distance} from Claim~\ref{claim:diff_priv_distance} and taking into account~\eqref{eq:worst_case_div}, we have that
\begin{equation}
	\KL{P_{W|S=s}}{Q_W} \leq \frac{1}{2} \mu^2 N^2,
	\label{eq:worst_case_div_gdp}
\end{equation}
which in combination with~\eqref{eq:prop4_prob_s} means that
\begin{equation}
	\bE_{s \sim P_S} \big[ \KL{P_{W|S=s}}{Q_W} \mid S \notin \cT \big] P_S(\cS \setminus \cT) \leq |\cZ| \mu^2.
	\label{eq:second_term_rhs_prop4_gdp}
\end{equation}

Taking into account Claim~\ref{claim:diff_priv_distance} and that we are considering only the covering of $\cT$, the first term on the r.h.s. of~\eqref{eq:prop4_total_exp} is now 
\begin{align}
\MoveEqLeft
\bE_{s \sim P_S} \big[ \KL{P_{W|S=s}}{Q_W} \mid S \in \cT \big] \nonumber\\
 &\leq \frac{1}{2} \frac{N \log N}{t^2} |\cZ|^2 \mu^2 +|\cZ| \log t,
 \label{eq:prop4_partial_exp_gdp}
\end{align}
which is minimized when $t = \sqrt{|\cZ| N \log N} \mu$.
If we operate analogously as for $\epsilon$-DP algorithms, we obtain~\eqref{eq:prop_mGDP_3_main_1} when $\mu \leq 2/\sqrt{|\cZ|}$, and~\eqref{eq:prop_mGDP_3_main_2} otherwise.
This concludes the proof of Proposition~\ref{prop:mi_mGDP_3}.
\hfill\IEEEQED

\subsection{Proof of Lemma~\ref{lemma:number_of_hypercubes_under_simplex}}
\label{app:proof_lemma_hypercubes}

\begin{figure}[t]
 \centering
 \includegraphics{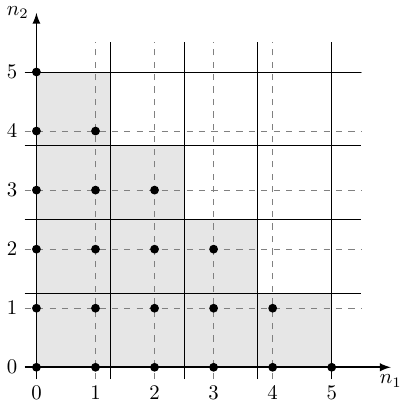}
 \caption{Example for $N=5$ samples, $K=2$ dimensions, and $t=4$ parts per dimension. Here, $n_1$ and $n_2$ are the dimensions of the $[0,5]^{2}$ hypercube. The number of highlighted blocks is $S_{2}(4)=10$. 
 \label{fig:hypercubes_simplex}}
\end{figure}

We start the proof by observing that, if the segment $[0,N]$ is divided in $t$ parts, we need all $t$ parts to cover it, i.e., $S_1(t) = t$.
Then, we note that the number of squares from a regular $t \times t$ grid needed to cover the area under the $1$-simplex is given by
\begin{equation}
    S_2(t) = \sum_{j=1}^t (t-j+1) = \frac{1}{2} t(t+1),
    \label{eq:hypercubes_simplex_1}
\end{equation}
since we need all vertical squares in the first column, and one less for every additional column, see Figure~\ref{fig:hypercubes_simplex} for a visual example.
We observe that it is possible to rewrite~\eqref{eq:hypercubes_simplex_1} as
\begin{equation}
    S_2(t) = \sum_{j=1}^t j = \sum_{j=1}^t S_1(j).
\end{equation}
This formulation is quite intuitive since the base of the pyramid has $S_1(t)$ blocks, the second level has $S_1(t-1)$ blocks, and so on, until we reach the level $t$ which has $S_1(1)$ blocks.
Furthermore, this recursion applies to any dimension, i.e.,
\begin{equation}
    S_K(t) = \sum_{j=1}^t S_{K-1}(j).
    \label{eq:hypercubes_simplex_2}
\end{equation}
For example, the number of cubes needed to cover the volume under the $2$-simplex is given by
\begin{equation}
    S_3(t) = \sum_{j=1}^t \frac{1}{2} j(j+1) = \frac{1}{6} t(t+1)(t+2).
\end{equation}

In what follows, we show that the recursion in~\eqref{eq:hypercubes_simplex_2} may be directly computed as $S_K(t)=f_K(t)$, where
\begin{equation}
 f_K(t) \triangleq \frac{1}{K!} \frac{(t+K-1)!}{(t-1)!}.
 \label{eq:hypercubes_simplex_3}
\end{equation} 
We prove this by induction.
We have already shown that $S_K(t)=f_K(t)$ for $K=1$, $2$, and $3$. Now, assuming it is true for $K-1$, we show it also holds for $K$.
First, we note the following equality

\begin{equation}
 f_{K-1}(t) = \begin{cases}
               f_K(t) - f_K(t-1) & \textnormal{if } t>1,\\
               1 & \textnormal{if } t=1,
              \end{cases}
 \label{eq:hypercubes_simplex_4}
\end{equation}
which may be obtained by some simple algebraic manipulations.
Then, assuming $t>1$,
\begin{align}
 S_K(t) &= \sum\nolimits_{j=1}^t S_{K-1}(j) \nonumber\\
 &\stack{a}{=} \sum\nolimits_{j=1}^t f_{K-1}(j) \nonumber\\
 &\stack{b}{=} 1+ \sum\nolimits_{j=2}^t f_K(j) - f_K(j-1) \nonumber\\
 &= f_K(t),
\end{align}
where $(a)$ comes from the induction assumption, and $(b)$ is due to~\eqref{eq:hypercubes_simplex_4}.
The case $t=1$ follows trivially since $f_K(1)=1$ for any $K \in \mathbb{Z}_+$; thus, the first part of~\eqref{eq:number_of_hypercubes_under_simplex} is proved.

The upper bound on $S_K(t)$, the second part of~\eqref{eq:number_of_hypercubes_under_simplex}, stems from the following small result.

\begin{claim}
\label{claim:factorials_to_exponential}
For any $t,M \in \mathbb{Z}_+$,
\begin{equation}
    \frac{(t+M)!}{(t-1)!} \leq \left(t+\frac{M}{2}\right)^{M+1}{.}
    \label{eq:claim_statement}
\end{equation}
\end{claim}
\begin{IEEEproof}
See Appendix~\ref{app:proof_factorials_to_exponential}{.}
\end{IEEEproof}

We may then proceed to bound the corresponding 
{factor} in~\eqref{eq:hypercubes_simplex_3} with~\eqref{eq:claim_statement}, where we set $M=K-1$. This concludes the proof of Lemma~\ref{lemma:number_of_hypercubes_under_simplex}.
\hfill\IEEEQED

\subsection{Proofs of secondary results}

\subsubsection{Proof of Claim~\ref{claim:num_types}}
\label{app:proof_claim_num_types}

The exact number of types is given by \cite[Problem~2.1]{csiszar_2011_information}
\begin{equation}
	|\cT_{\cZ,N}| = \binom{N+|\cZ|-1}{|\cZ|-1}.
\end{equation}
Then, we may bound $|\cT_{\cZ,N}|$ from above as follows,
\begin{align}
	|\cT_{\cZ,N}| &= \frac{(N+|\cZ|-1)!}{(|\cZ|-1)!\,N!} \nonumber \\
	&= \prod\nolimits_{i=1}^{|\cZ|-1} \left(1 + \frac{N}{i} \right) \nonumber \\
	&\leq \prod\nolimits_{i=1}^{|\cZ|-1} (1 + N) \nonumber \\
	&= (N+1)^{|\cZ|-1},
\end{align}
with equality if and only if $|\cZ| = 2$.
\hfill\IEEEQED

\subsubsection{Proof of Claim~\ref{claim:factorials_to_exponential}}
\label{app:proof_factorials_to_exponential}

We start the proof by noting the following upper bound, that is valid for any $t,a,b\in\bR$:
\begin{equation}
    (t+a)(t+b)\leq \left(t +\frac{a+b}{2}\right)^{2},
    \label{eq:bound_prod_center}
\end{equation}
which is a simple consequence of the equality
\begin{equation*}
    (t+a)(t+b)+\frac{(a-b)^2}{4} = \left(t +\frac{a+b}{2}\right)^{2}.
\end{equation*}

Therefore, if $M = 2L$, $L \in \mathbb{Z}_+$, there is an odd number of 
{factors} ($2L+1$) on the l.h.s. of~\eqref{eq:claim_statement}. We may pair all the 
{factors} that are equidistant from the center, with the exception of the middle one, and bound them using~\eqref{eq:bound_prod_center} to obtain
\begin{equation*}
    (t+L)^{2L}(t+L) = \left(t+\frac{M}{2}\right)^{M+1}.
\end{equation*}
On the other hand, if $M = 2L + 1$, $L \in \mathbb{Z}_+$, there is now an even number of 
{factors} ($2L+2$) on the l.h.s. of~\eqref{eq:claim_statement}. 
We may nonetheless pair all the 
{factors} that are equidistant from the center and, by using~\eqref{eq:bound_prod_center}, also obtain the bound
\begin{equation*}
    \left(t + \frac{2L +1}{2}\right)^{2L+2} = \left(t+\frac{M}{2}\right)^{M+1}.\IEEEQEDhereeqn
\end{equation*}

\section*{Acknowledgment}

The authors are grateful to the Associate Editor and to the anonymous reviewers for their constructive and helpful comments on the earlier version of the paper, which helped us improve the manuscript. The authors would also like to thank Mario Diaz for pointing out~\cite{kaji_bounds_2015}, which lead to Remark~\ref{rem:multinomial}.

\bibliographystyle{IEEEtran}
\bibliography{IEEEabrv,references}

\begin{thebibliography}{10}
\providecommand{\url}[1]{#1}
\csname url@samestyle\endcsname
\providecommand{\newblock}{\relax}
\providecommand{\bibinfo}[2]{#2}
\providecommand{\BIBentrySTDinterwordspacing}{\spaceskip=0pt\relax}
\providecommand{\BIBentryALTinterwordstretchfactor}{4}
\providecommand{\BIBentryALTinterwordspacing}{\spaceskip=\fontdimen2\font plus
\BIBentryALTinterwordstretchfactor\fontdimen3\font minus
  \fontdimen4\font\relax}
\providecommand{\BIBforeignlanguage}[2]{{%
\expandafter\ifx\csname l@#1\endcsname\relax
\typeout{** WARNING: IEEEtran.bst: No hyphenation pattern has been}%
\typeout{** loaded for the language `#1'. Using the pattern for}%
\typeout{** the default language instead.}%
\else
\language=\csname l@#1\endcsname
\fi
#2}}
\providecommand{\BIBdecl}{\relax}
\BIBdecl

\bibitem{stoica1989maximum}
P.~Stoica, R.~L. Moses, B.~Friedlander, and T.~Soderstrom, ``Maximum likelihood
  estimation of the parameters of multiple sinusoids from noisy measurements,''
  \emph{IEEE Transactions on Acoustics, Speech, and Signal Processing},
  vol.~37, no.~3, pp. 378--392, Mar. 1989.

\bibitem{havaei2017brain}
M.~Havaei, A.~Davy, D.~Warde-Farley, A.~Biard, A.~Courville, Y.~Bengio, C.~Pal,
  P.-M. Jodoin, and H.~Larochelle, ``Brain tumor segmentation with deep neural
  networks,'' \emph{Medical image analysis}, vol.~35, pp. 18--31, 2017.

\bibitem{russo_2020_bias}
D.~Russo and J.~Zou, ``How much does your data exploration overfit?
  {C}ontrolling bias via information usage,'' \emph{IEEE Transactions on
  Information Theory}, vol.~66, no.~1, pp. 302--323, Jan. 2020.

\bibitem{xu2017information}
A.~Xu and M.~Raginsky, ``Information-theoretic analysis of generalization
  capability of learning algorithms,'' in \emph{31st Conference on Neural
  Information Processing Systems (NeurIPS)}, Dec. 2017, pp. 2524--2533.

\bibitem{jiao_2017_dependence}
J.~Jiao, Y.~Han, and T.~Weissman, ``Dependence measures bounding the
  exploration bias for general measurements,'' in \emph{2017 IEEE International
  Symposium on Information Theory (ISIT)}, Jun. 2017, pp. 1475--1479.

\bibitem{bu_2019_tightening}
Y.~Bu, S.~Zou, and V.~V. Veeravalli, ``Tightening mutual information based
  bounds on generalization error,'' in \emph{2019 IEEE International Symposium
  on Information Theory (ISIT)}, Jul. 2019, pp. 587--591.

\bibitem{dwork_generalization_2015}
C.~Dwork, V.~Feldman, M.~Hardt, T.~Pitassi, O.~Reingold, and A.~Roth,
  ``Generalization in adaptive data analysis and holdout reuse,'' in \emph{29th
  Conference on Neural Information Processing Systems (NeurIPS)}, Dec. 2015,
  pp. 2350--2358.

\bibitem{dwork_preserving_2015}
C.~Dwork, V.~Feldman, M.~Hardt, T.~Pitassi, O.~Reingold, and A.~L. Roth,
  ``Preserving statistical validity in adaptive data analysis,'' in \emph{47th
  Annual ACM Symposium on Theory of Computing (STOC)}, Jun. 2015, pp. 117--126.

\bibitem{dwork_dp_2014}
C.~Dwork and A.~Roth, ``{The Algorithmic Foundations of Differential
  Privacy},'' \emph{Foundations and Trends{\textregistered} in Theoretical
  Computer Science}, vol.~9, no. 3--4, pp. 211--407, 2014.

\bibitem{dong2019gaussian}
J.~Dong, A.~Roth, and W.~Su, ``Gaussian differential privacy,'' \emph{Journal
  of the Royal Statistical Society}, 2021.

\bibitem{shalev2014understanding}
S.~Shalev-Shwartz and S.~Ben-David, \emph{{Understanding machine learning: From
  theory to algorithms}}.\hskip 1em plus 0.5em minus 0.4em\relax Cambridge
  university press, 2014.

\bibitem{asadi2018chaining}
A.~Asadi, E.~Abbe, and S.~Verd{\'u}, ``Chaining mutual information and
  tightening generalization bounds,'' in \emph{32nd Conference on Neural
  Information Processing Systems (NeurIPS)}, Dec. 2018, pp. 7234--7243.

\bibitem{issa2019strengthened}
I.~Issa, A.~R. Esposito, and M.~Gastpar, ``Strengthened information-theoretic
  bounds on the generalization error,'' in \emph{2019 IEEE International
  Symposium on Information Theory (ISIT)}, Jul. 2019, pp. 582--586.

\bibitem{wang2019information}
H.~Wang, M.~Diaz, J.~C.~S. {Santos Filho}, and F.~P. Calmon, ``An
  information-theoretic view of generalization via {W}asserstein distance,'' in
  \emph{2019 IEEE International Symposium on Information Theory (ISIT)}, Jul.
  2019, pp. 577--581.

\bibitem{bousquet2002stability}
O.~Bousquet and A.~Elisseeff, ``Stability and generalization,'' \emph{Journal
  of machine learning research}, vol.~2, no. Mar, pp. 499--526, 2002.

\bibitem{raginsky2016information}
M.~Raginsky, A.~Rakhlin, M.~Tsao, Y.~Wu, and A.~Xu, ``Information-theoretic
  analysis of stability and bias of learning algorithms,'' in \emph{2016 IEEE
  Information Theory Workshop (ITW)}, Sep. 2016, pp. 26--30.

\bibitem{bassily_algorithmic_2016}
R.~Bassily, K.~Nissim, A.~Smith, T.~Steinke, U.~Stemmer, and J.~Ullman,
  ``Algorithmic stability for adaptive data analysis,'' in \emph{48th Annual
  ACM Symposium on Theory of Computing (STOC)}, Jun. 2016, pp. 1046--1059.

\bibitem{feldman2019high}
V.~Feldman and J.~Vondrak, ``High probability generalization bounds for
  uniformly stable algorithms with nearly optimal rate,'' in \emph{Conference
  on Learning Theory (COLT)}.\hskip 1em plus 0.5em minus 0.4em\relax PMLR,
  2019, pp. 1270--1279.

\bibitem{jung_new_2020}
C.~Jung, K.~Ligett, S.~Neel, A.~Roth, S.~Sharifi-Malvajerdi, and M.~Shenfeld,
  ``{A New Analysis of Differential Privacy's Generalization Guarantees},'' in
  \emph{11th {Innovations} in {Theoretical} {Computer} {Science} {Conference}
  ({ITCS} 2020)}, vol. 151, 2020, pp. 31:1--31:17.

\bibitem{buldygin1980sub}
V.~V. Buldygin and Y.~V. Kozachenko, ``Sub-gaussian random variables,''
  \emph{Ukrainian Mathematical Journal}, vol.~32, no.~6, pp. 483--489, 1980.

\bibitem{polyanskiy2014lecture}
\BIBentryALTinterwordspacing
Y.~Polyanskiy and Y.~Wu, ``{Lecture notes on Information Theory},'' 2019.
  [Online]. Available:
  \url{http://www.stat.yale.edu/~yw562/teaching/itlectures.pdf}
\BIBentrySTDinterwordspacing

\bibitem{hellstrom2020generalization}
F.~Hellstr{\"o}m and G.~Durisi, ``Generalization bounds via information density
  and conditional information density,'' \emph{IEEE Journal on Selected Areas
  in Information Theory}, vol.~1, no.~3, pp. 824--839, 2020.

\bibitem{cover2012elements}
T.~M. Cover and J.~A. Thomas, \emph{{Elements of Information Theory}},
  2nd~ed.\hskip 1em plus 0.5em minus 0.4em\relax John Wiley \& Sons, 2006.

\bibitem{hershey_2007_approx}
J.~R. {Hershey} and P.~A. {Olsen}, ``Approximating the {K}ullback {L}eibler
  divergence between {G}aussian mixture models,'' in \emph{2007 IEEE
  International Conference on Acoustics, Speech, and Signal Processing
  (ICASSP)}, vol.~4, Apr. 2007, pp. 317--320.

\bibitem{kaji_bounds_2015}
Y.~{Kaji}, ``Bounds on the entropy of multinomial distribution,'' in \emph{IEEE
  International Symposium on Information Theory (ISIT)}, Jun. 2015, pp.
  1362--1366.

\bibitem{yu2014differentially}
F.~Yu, M.~Rybar, C.~Uhler, and S.~E. Fienberg, ``Differentially-private
  logistic regression for detecting multiple-snp association in gwas
  databases,'' in \emph{International Conference on Privacy in Statistical
  Databases}.\hskip 1em plus 0.5em minus 0.4em\relax Springer, 2014, pp.
  170--184.

\bibitem{wainwright2019high}
M.~J. Wainwright, \emph{High-dimensional statistics: A non-asymptotic
  viewpoint}.\hskip 1em plus 0.5em minus 0.4em\relax Cambridge University
  Press, 2019, vol.~48.

\bibitem{mironov_renyi_2017}
I.~Mironov, ``Rényi {Differential} {Privacy},'' in \emph{2017 {IEEE} 30th
  {Computer} {Security} {Foundations} {Symposium} ({CSF})}, Aug. 2017, pp.
  263--275.

\bibitem{bu2019deep}
Z.~Bu, J.~Dong, Q.~Long, and W.~J. Su, ``Deep learning with gaussian
  differential privacy,'' \emph{Harvard data science review}, vol. 2020,
  no.~23, 2020.

\bibitem{kingma2014adam}
D.~P. Kingma and J.~Ba, ``{Adam: A method for stochastic optimization},''
  \emph{arXiv preprint arXiv:1412.6980}, 2014.

\bibitem{heikkila2019differentially}
M.~Heikkil{\"a}, J.~J{\"a}lk{\"o}, O.~Dikmen, and A.~Honkela, ``Differentially
  private markov chain monte carlo,'' in \emph{Advances in Neural Information
  Processing Systems (NeurIPS)}, 2019, pp. 4113--4123.

\bibitem{rubinstein2009learning}
B.~I. Rubinstein, P.~L. Bartlett, L.~Huang, and N.~Taft, ``Learning in a large
  function space: Privacy-preserving mechanisms for svm learning,''
  \emph{Journal of Privacy and Confidentiality}, vol.~4, no.~1, pp. 65--100,
  2012.

\bibitem{steinke2020reasoning}
T.~Steinke and L.~Zakynthinou, ``Reasoning about generalization via conditional
  mutual information,'' in \emph{Conference on Learning Theory (COLT)}.\hskip
  1em plus 0.5em minus 0.4em\relax PMLR, 2020, pp. 3437--3452.

\bibitem{haghifam2020sharpened}
M.~Haghifam, J.~Negrea, A.~Khisti, D.~M. Roy, and G.~K. Dziugaite, ``Sharpened
  generalization bounds based on conditional mutual information and an
  application to noisy, iterative algorithms,'' \emph{Advances in Neural
  Information Processing Systems (NeurIPS)}, vol.~33, pp. 9925--9935, 2020.

\bibitem{rodriguez2020random}
B.~Rodr{\'\i}guez-G{\'a}lvez, G.~Bassi, R.~Thobaben, and M.~Skoglund, ``On
  random subset generalization error bounds and the stochastic gradient
  langevin dynamics algorithm,'' in \emph{2020 IEEE Information Theory Workshop
  (ITW)}.\hskip 1em plus 0.5em minus 0.4em\relax IEEE, 2021, pp. 1--5.

\bibitem{JMLR:v17:15-313}
Y.-X. Wang, J.~Lei, and S.~E. Fienberg, ``Learning with differential privacy:
  Stability, learnability and the sufficiency and necessity of erm principle,''
  \emph{Journal of Machine Learning Research}, vol.~17, no. 183, pp. 1--40,
  2016.

\bibitem{bun2016concentrated}
M.~Bun and T.~Steinke, ``Concentrated differential privacy: Simplifications,
  extensions, and lower bounds,'' in \emph{Theory of Cryptography
  Conference}.\hskip 1em plus 0.5em minus 0.4em\relax Springer, 2016, pp.
  635--658.

\bibitem{5992163}
G.~{Barthe} and B.~{K{\"o}pf}, ``Information-theoretic bounds for
  differentially private mechanisms,'' in \emph{2011 IEEE 24th Computer
  Security Foundations Symposium}, 2011, pp. 191--204.

\bibitem{mcdonald1999course}
J.~N. McDonald and N.~A. Weiss, \emph{{A Course in Real Analysis}},
  2nd~ed.\hskip 1em plus 0.5em minus 0.4em\relax Cambridge, Massachusetts:
  Elsevier, 2013.

\bibitem{csiszar_2011_information}
I.~Csisz{\'a}r and J.~K{\"o}rner, \emph{{Information Theory: Coding Theorems
  for Discrete Memoryless Systems}}, 2nd~ed.\hskip 1em plus 0.5em minus
  0.4em\relax Cambridge University Press, 2011.

\end{thebibliography}

\begin{IEEEbiographynophoto}{Borja Rodr\'iguez-G\'alvez}
(S'20) was born in Girona, Catalonia, Spain, in 1995. He received the B.Sc. in Telecommunications Engineering from the Universitat Polit\`ecnica de Catalunya, Spain, in 2017 and the M.Sc. in Machine Learning from KTH Royal Institute of Technology, Sweden, in 2019. He is currently pursuing a Ph.D. in Electrical Engineering at KTH Royal Institute of Technology, Sweden.

Mr. Rodr\'iguez-G\'alvez is a student member of the IEEE Information Theory Society. His research interests span information theory, probability theory, and statistics, with applications to machine learning, privacy, and fairness.
\end{IEEEbiographynophoto}

\begin{IEEEbiographynophoto}{Germ\'an Bassi}
(S'10--M'16) received the B.Sc.\ and M.Sc.\ degrees in Electrical Engineering from
the University of Buenos Aires, Argentina, in 2010, and the Ph.D.\ degree in Telecommunications
from CentraleSup{\'e}lec, France, in 2015. From 2016 to 2020, he was a postdoctoral researcher at
KTH Royal Institute of Technology, Sweden. 

Dr.~Bassi is now a researcher at Ericsson Research, Stockholm, Sweden, where he works in next-generation communication technologies.
His research interests include wireless communications, network information theory, physical-layer security, and inference and statistics, with applications to privacy and machine learning.
\end{IEEEbiographynophoto}

\begin{IEEEbiographynophoto}{Mikael Skoglund}
(S'93--M'97--SM'04--F'19) received the Ph.D.~degree in
1997 from Chalmers University of Technology, Sweden.  In 1997, he joined the Royal Institute of Technology (KTH), Stockholm, Sweden, where he was appointed to the Chair in Communication Theory in 2003.  At KTH, he heads the Division of Information Science and Engineering, and the
Department of Intelligent Systems. 

Dr.~Skoglund has worked on problems in source-channel coding, coding and transmission for wireless communications, Shannon theory, information and control, and statistical signal processing. He has authored and co-authored more than 180 journal and 400 conference papers.

Dr.~Skoglund is a Fellow of the IEEE. During 2003--08 he was an associate editor for the IEEE Transactions on Communications. During 2008--12 he was on the editorial board for the IEEE Transactions on Information Theory and he joined it again in 2021. He has served on numerous technical program committees for IEEE sponsored conferences, he was general co-chair for IEEE ITW 2019, and is serving as TPC co-chair for IEEE ISIT 2022.
\end{IEEEbiographynophoto}

\end{document}